\documentclass[useAMS]{mn2e}

\usepackage{amssymb,amsmath,psfig,times}
\def\gsim{ \lower .75ex \hbox{$\sim$} \llap{\raise .27ex \hbox{$>$}} }
\def\lsim{ \lower .75ex\hbox{$\sim$} \llap{\raise .27ex \hbox{$<$}} }

\def\sc{Schwarzschild}

\title[Jet and accretion power in the most powerful Fermi blazars]
{Jet and accretion power in the most powerful Fermi blazars}  
\author[G. Ghisellini,  F. Tavecchio and G. Ghirlanda]
{G. Ghisellini\thanks{Email:
gabriele.ghisellini@brera.inaf.it}, F. Tavecchio and G. Ghirlanda
\\
INAF -- Osservatorio Astronomico di Brera, Via Bianchi 46, I--23807 Merate, Italy\\
}

\begin{document}  

\maketitle

\begin{abstract}
Among the blazars detected by the {\it Fermi} satellite, we have selected
the 23 blazars that in the three months of survey had an average
$\gamma$--ray luminosity above $10^{48}$ erg s$^{-1}$.
For 17 out of the 23 sources we found and analysed X--ray and optical--UV data
taken by the {\it Swift} satellite.
With these data, implemented by archival and not simultaneous data,
we construct the spectral energy distributions, and interpreted them
with a simple one--zone, leptonic, synchrotron and inverse Compton model.
When possible, we also compare different high energy states
of single sources, like 0528+134 and 3C 454.3, for which multiple good sets
of multi--wavelength data are available.
In our powerful blazars the high energy emission
always dominates the electromagnetic output, and the
relatively low level of the synchrotron radiation often does not
hide the accretion disk emission.
We can then constrain the black hole mass and the disk luminosity. 
Both are large (i.e. masses equal or greater than $10^9 M_\odot$ and 
disk luminosities above 10\% of Eddington).
By modelling the non--thermal continuum we derive the power that the
jet carries in the form of bulk motion of particles and fields.
On average, the jet power is found to be slightly larger than the disk luminosity,
and proportional to the mass accretion rate. 
\end{abstract}

\section{Introduction}

We would like to attack in a more systematic way than
done in the past the problem of the relation between 
the power carried by the jet of blazars in the form of bulk motion 
of particles and magnetic fields, and of the luminosity associated to accretion.
This task is not easy, mainly because of the Doppler enhancement of 
the non--thermal continuum produced by jets, and because 
this continuum very often hides the radiation produced by the accretion disk.
But starting from the early work of Rawlings \& Saunders (1991),
evidence has been accumulated that the jet power of
extragalactic radio--loud sources is at least
of the same order as the accretion one, and possibly a factor $\sim$10 larger.

On the larger scale (radio--lobe size) one can resort to minimum
energy considerations and estimates of the lifetime of radio--lobes
to find out the minimum jet power needed to sustain
the radio--lobe emission (Rawlings \& Saunders 1991).
The main uncertainty associated to this argument is the 
unknown energy contained in the proton component.

At smaller, but still very large, jet scales 
(on kpc to Mpc), one can use
the recently discovered (by the {\it Chandra} satellite) X--ray emission 
from resolved knots, to model it and to infer the total number of leptons needed 
to produce the observed radiation. 
Here the main uncertainties are related to the number of low energy
leptons (of energy $\gamma_{\rm min}m_{\rm e}c^2$
present in the source, a number that is rarely well constrained.
Furthermore, one also needs to assume how many protons are associated
for each emitting lepton.
Assuming a one to one correspondence and using observed
data to limit $\gamma_{\rm min}$, Tavecchio et al. (2000); Celotti, Ghisellini 
\& Chiaberge (2001), Tavecchio et al. (2004), Sambruna et al. (2006) and 
Tavecchio et al. (2007) derived jet powers that are comparable with those 
inferred at the sub--pc scales, suggesting that the jet powers 
in these sources are conserved all along the jet,
with the caveat that for large scale jets is difficult
to constrain the bulk Lorentz factor $\Gamma$.
However, values compatible with the inner jet
are preferred (see Ghisellini \& Celotti 2001).

A technique developed quite recently makes use of the cavities or ``bubbles"
in the X--ray emitting intra--cluster medium of cluster of galaxies,
and measures the energy required to inflate such bubbles.
Assuming that this energy is furnished by the jet, one can calculate
the associated jet power (i.e. Churazov et al. 2002; Allen et al. 2006;
Balmaverde, Balbi \& Capetti 2008).

At the VLBI scale (pc or tens of pc), one takes advantage of the
resolving power of VLBI to measure the size of the synchrotron emitting
region: by requiring that the Self Compton process does not
overproduce the total observed X--ray flux, one derives a lower
limit to the Doppler factor $\delta$ [defined
as $\delta=1/[\Gamma(1-\beta\cos\theta_{\rm v})]$, 
where $\theta_{\rm v}$ is the viewing angle and $\Gamma$ the bulk Lorentz factor],
and a limit on the number of emitting particles and 
on the value of the magnetic field (Celotti \& Fabian 1993).
This in turn gives an estimate of the jet power.
The main uncertainties here are the number of low energy electrons
(they emit unobservable self--absorbed synchrotron emission)
and the fraction of the total X--ray flux produced by the radio VLBI knot.

The advent of the {\it Compton Gamma Ray Observatory} and its
high energy EGRET instrument on one side, and of ground based
Cherenkov telescopes on the other, finally let it possible
to discover at what frequencies blazar jets emit most of 
their power, i.e. above MeV energies.
Variability of the high energy flux also tells us that the 
emission site cannot be too distant from the black hole,
while the absence of $\gamma$--$\gamma$ absorption (leading to pair
production) tells us that the emission site cannot be too close
to the black hole and its accretion disk (Ghisellini \& Madau 1996;
Ghisellini \& Tavecchio 2009; hereafter GT09).
Bracketed by these two limits, one obtains a few hundreds of \sc\
radii as the preferred jet location where most of the dissipation occurs.
Modelling the observed spectral energy distribution (SED) from the mm to the
$\gamma$--ray band returns the particle number and the field strength
needed to account for the observed data (see e.g. 
Celotti \& Ghisellini 2008, hereafter CG08; Kataoka et al. 2008, 
Maraschi \& Tavecchio 2003).
Again, the main uncertainty in the game is $\gamma_{\rm min}$,
but in some cases $\gamma_{\rm min}$ can be strongly constrained.
This occurs when the considered blazar has good soft X--ray data
characterised by a flat spectrum.
In these cases the X--ray emission is likely to be due to 
inverse Compton scatterings between low energy electrons and 
seed photons produced externally to the jet (external Compton, EC;
e.g. Sikora, Begelman \& Rees 1994).
In fact, while the synchrotron Self--Compton (SSC) process
uses relatively high energy electrons and internal synchrotron
radiation to produce X--rays, in the EC case we do see X--rays
produced by electrons of very low energies, strongly constraining 
$\gamma_{\rm min}$  (see e.g. Tavecchio et al. 2007).

As for the accretion disk radiation, it is usually
hidden by the stronger optical--UV synchrotron emission.
But in very powerful blazars we have good reasons to believe that the synchrotron
component originating the radio to IR--optical radiation is not completely hiding
the thermal emission produced by the accretion disk.
These reasons come partly from past observations of relatively high redshifts
blazars (even if not detected by {\it Fermi} or EGRET, see e.g. Landt et al.
2008; Maraschi et al. 2008; Sambruna et al. 2007; GT09) 
and partly from theoretical considerations about the expected scaling of the 
magnetic energy density with the black hole mass:
we expect lower magnetic fields in jets associated to 
larger black hole masses (e.g. GT09),
and thus a reduced importance of the synchrotron component of the SED.
Furthermore, the presence, in these objects, of broad emission lines of
relatively ``normal" equivalent width (compared to radio--quiet objects)
ensures that the thermal ionising continuum cannot be much below the observed
optical emission.

The above considerations guide us to select the most powerful blazars
as very good candidates for measuring both the jet power and the accretion luminosity.
In turn, since in these sources most of the luminosity is emitted 
in the hard X--rays and in the $\gamma$--ray band, it is natural to take advantage
of the recently published list of blazars detected by the 
Large Area Telescope (LAT) on board the {\it Fermi Gamma Ray Space Telescope (Fermi)}.
It revealed more than one hundred blazars with a significance larger than 10$\sigma$
in the first three months of operation (Abdo et al. 2009a, hereafter A09).
Of these, 57 are classified as flat spectrum radio quasars (FSRQs), 
42 as BL Lac objects, while for 5 sources the classification is uncertain. 
Including  2 radio--galaxies the total number of extragalactic sources amounts to 106.
Redshifts are known for all FSRQs, for 30 BL Lacs, for 1 source
of uncertain classification and for the two radio--galaxies,
for a total of 90 objects.
Within this sources, there are 23 blazars exceeding an average $\gamma$--ray luminosity
of $10^{48}$ erg s$^{-1}$ within the 3 months survey. 
These are the targets of our study, aimed to estimate the power 
associated to accretion and to the jet.
We will also take advantage of the optical--UV and X--ray observations
made by the {\it Swift} satellite, to construct the spectral energy distributions
(SED) of {\it Fermi} blazars, a crucial ingredient to reliably constrain 
the emission models leading to the jet and accretion power estimate.

In this paper we use a cosmology with $h=\Omega_\Lambda=0.7$ and $\Omega_{\rm M}=0.3$,
and use the notation $Q=Q_X 10^X$ in cgs units (except for the black hole masses,
measured in solar mass units).

\begin{table}
\centering
\begin{tabular}{lllllll}
\hline
\hline
Name  &Alias &$z$ &$\log L_\gamma$ &E? &{\it S?} &LC?\\
\hline   
PKS 0048--071 &               &1.975  &48.2  &  &   &  \\
PKS 0202--17  &               &1.74   &48.2  &  &   &   \\
PKS 0215+015  &               &1.715  &48.16 &  &Y  &  \\
PKS 0227--369 &               &2.115  &48.6  &  &Y  &  \\
AO  0235+164  &               &0.94   &48.4  &Y &Y  &Y  \\
PKS 0347--211 &               &2.944  &49.1  &  &Y  & \\
PKS 0426--380 &               &1.112  &48.06 &  &Y  &  \\
PKS 0454--234 &               &1.003  &48.16 &Y &Y  &Y \\
PKS 0528+134  &               &2.04   &48.8  &Y &Y  &Y \\
TXS 0820+560  &S4 0820+56     &1.417  &48.005&  &Y  &  \\
TXS 0917+449  &RGB J0920+446  &2.1899 &48.4  &Y &Y  &  \\
TXS 1013+054  &PMN J1016+051  &1.713  &48.2  &  &   &  \\
PKS 1329--049 &               &2.15   &48.5  &  &   &  \\
PKS 1454--354 &               &1.424  &48.5  &Y &Y  &Y \\
PKS 1502+106  &               &1.839  &49.1  &  &Y  &Y \\
TXS 1520+319  &B2 1520+31     &1.487  &48.4  &  &Y  &Y \\
PKS 1551+130  &               &1.308  &48.04 &  &   &  \\
TXS 1633+382  &4C +38.41      &1.814  &48.6  &Y &Y  &Y  \\
PKS 2023--077 &               &1.388  &48.6  &Y &Y  &  \\
PKS 2052-47   &               &1.4910 &48.03 &Y &   &  \\
PKS 2227--088 &PHL 5225       &1.5595 &48.2  &  &Y  &  \\
PKS 2251+158  &3C 454.3       &0.859  &48.7  &Y &Y  &Y \\
PKS 2325+093  &               &1.843  &48.5  &  &Y  &  \\
\hline
\hline 
\end{tabular}
\vskip 0.4 true cm
\caption{The 23 most powerful {\it Fermi} blazars in the A09 catalogue.
In the last 3 columns we indicate the logarithm of the average $\gamma$--ray
luminosity as observed by {\it Fermi} during the first 3 months of survey
(cgs units), if the source was detected by EGRET, if there are {\it Swift}
observations, and if there is the public available $\gamma$--ray light--curve
at the web page:
{\tt http://fermi.gsfc.nasa.gov/ssc/data/access/lat/} .
}
\label{sample}
\end{table}

\section{The sample}

Tab. \ref{sample} lists 23 blazars taken from the
A09 catalogue of {\it Fermi} detected blazars.
We selected them simply on the basis of their K--corrected 
(see e.g. Ghisellini, Maraschi \& Tavecchio 2009) $\gamma$--ray luminosity 
being larger than $10^{48}$ erg s$^{-1}$.
All sources but two are Flat Spectrum Radio Quasars (FSRQs).
The exceptions are AO 0235+164 and PKS 0426--380, 
classified as BL Lac objects. 
However, these sources do have broad emission lines, 
(see e.g. Raiteri et al. 2007 for AO 0235+164 and Sbarufatti et al.
2005 for PKS 0426--380), visible in low emission states.
We therefore believe that AO 0235+164 and PKS 0426--380 are FSRQs 
whose line emission is often swamped by the enhanced 
non--thermal continuum.
In Tab. \ref{sample}
we mark the 8 blazars for which there are public available {\it Fermi} light curves.
Although we use for all blazars the 3--months averaged $\gamma$--ray flux and
spectral index, for these 8 sources we can check if at the time of the {\it Swift}
observations the $\gamma$--ray integrated photon flux was much different than the average
(the spectral index is available only for the 3--months average).
For all sources but PKS 1502+106 we find a good consistency.
For PKS 1502+106 the {\it Swift} observations were probably performed
during the rapid decay phase after a major $\gamma$--ray flare.
Therefore it is difficult
to assess the exact $\gamma$--ray flux during the {\it Swift} observation,
but in any case it should not be different than the used average flux by more
than a factor 2.

\section{Swift observations and analysis}

For 17 of the 23 blazars in our sample there are {\it Swift} observations,
with the majority of the objects observed during
the 3 months of the {\it Fermi} survey.
The data were analysed with the 
most recent software \texttt{Swift\_Rel3.2} released as part of the \texttt{Heasoft v. 6.6.2}.
The calibration database is that updated to April 10, 2009. 
The XRT data were processed with the standard procedures ({\texttt{XRTPIPELINE v.0.12.2}). 
We considered photon counting (PC) mode data with the standard 0--12 grade selection. Source 
events were extracted in a circular region of  aperture $\sim 47''$, and background was estimated 
in a same sized circular region far from the source. 
Ancillary response files were created through the \texttt{xrtmkarf} task. 
The channels with energies below 0.2 keV and above 10 keV were excluded 
from the fit and the spectra were rebinned in energy so to have at 
least 30 counts per bin. 
Each spectrum was analysed through XSPEC
with an absorbed power--law with a fixed 
Galactic column density from Kalberla et al. (2005).
The computed errors represent the 90\% confidence interval on the spectral parameters.
Tab. \ref{xrt} reports the log of the observations and the results of
the fitting the X--ray data with a simple power law model.

UVOT (Roming et al. 2005) source counts were extracted from 
a circular region $5''-$sized centred on the source position, 
while the background was extracted from 
a larger circular nearby source--free region.
Data were integrated with the \texttt{uvotimsum} task and then 
analysed by using the  \texttt{uvotsource} task.
The observed magnitudes have been dereddened according to the formulae 
by Cardelli et al. (1989) and converted into fluxes by using standard 
formulae and zero points from Poole et al. (2008).
Tab. \ref{uvot} list the observed magnitudes in the 6 filters of UVOT,
and the Galactic extinction appropriate for each source.


%
\begin{table*}
\centering
\begin{tabular}{llllllll}
\hline
\hline
source        &OBS date    &$N^{\rm Gal}_{\rm H}$   &$\Gamma$       
                                           &$\chi2{\rm /dof}$ & $F_{\rm 0.2-10, unabs}$  &$F_{\rm 2-10, unabs}$ \\
                &dd/mm/yyyy     &$10^{20}$ cm$^{-2}$& & &10$^{-12}$  cgs &10$^{-12}$ cgs  \\
\hline
0215+015       &28/06/2005   &3.26    &1.48$\pm$0.1     &17/15   &3.1$\pm$0.3   &2.0$\pm$0.2\\
0227--369$^a$  &07/11/2008   &2.42    &1.22$\pm$0.14    &8.8/7   &1.6$\pm$0.2   &1.22$\pm$0.15 \\
0235+164       &02/09/2008   &7.7     &1.44$\pm$0.12    &20/14   &4.8$\pm $0.3  &3.6$\pm$0.3    \\
0347--211      &15/10/2008   &4.2     &1.55$\pm$0.47    &2.4/3   &1.0$\pm$0.25  &0.7$\pm$0.3    \\
0426--380      &27/10/2008   &2.1     &1.93$\pm$0.21    &1/3     &1.5$\pm$0.2   &0.7$\pm$0.2    \\
0454-234       &26/10/2008   &2.8     &1.6$\pm$0.3      &1/1     &1.2$\pm$0.2   &0.8$\pm$0.2    \\
0528+134       &02/10/2008   &24.2    &1.3$\pm$0.3      &1.8/4   &4.0$\pm$0.6   &3.0$\pm$0.5    \\
0820+560       &20/12/2008   &4.71    &1.65$\pm$0.16    &1.88/6  &1.34$\pm$0.16 &0.8$\pm$0.2 \\
0920+44$^a$    &18/01/2009   &1.47    &1.74$\pm$0.13    &13.4/12 &4.2$\pm$0.4   &2.6$\pm$0.2  \\
1454--354$_1$  &07/01/2008   &6.54    &1.86$\pm$0.24    &3.8/4   &1.3$\pm$0.3   &0.6$\pm$0.12 \\
1454--354$_2$  &12/09/2008   &6.54    &1.84$\pm$0.31    &1.06/1  &3.7$\pm$0.3   &1.8$\pm$0.4 \\
1502+106       &08/08/2008   &2.29    &1.55$\pm$0.10    &18.4/18 &2.6$\pm $0.2  &1.6$\pm$0.2  \\
1520+319       &12/11/2008   &2.0(*)  &1.17$\pm$0.4     &43/44   &0.6           &...        \\
1633+382       &22/02/2007   &1.1(*)  &1.56$\pm$0.25    &7.7/5   &1.58          &...           \\
2023--07       &08/12/2008   &3.37    &1.78$\pm$0.3     &3.8/3   &1.5$\pm$0.3   &0.7$\pm$0.3    \\
2227--088      &28/04/2005   &4.21    &1.55$\pm$0.12    &29.7/22 &4.5$\pm$0.4   &2.7$\pm$0.1    \\
2251+158       &08/08/2008   &6.58    &1.7$\pm$0.06     &485/527 &33$\pm$1      &518$ \pm$1    \\
2325+093       &03/06/2008   &4.1     &1.11$\pm$0.17    &6/5     &3.7$\pm$0.6   &2.9$ \pm$ 0.5    \\

\hline
\hline
\end{tabular}
\vskip 0.4 true cm
\caption{Results of the X--ray analysis.
*: poorly determined spectrum, the C--Statistic was used to fit the  spectrum. {\it a:}the spectrum was fitted from 0.5 keV. }
\label{xrt}
\end{table*}
\begin{table*}
\centering
\begin{tabular}{lllllllll}
\hline
\hline
source        &OBS date    &$V$             &$B$             &$U$              &$W1$            & $M2$           &$W2$            &$A_V$  \\
\hline
0215+015      &28/06/2008   & ...           &...             & ...             & ...            &...             &18.3$\pm$0.14   &0.111  \\
0227--369     &07/11/2008   &18.96$\pm$0.08 &19.42$\pm$0.05  &18.56$\pm $0.04  &18.94$\pm$0.05  &18.98$\pm$0.06  &18.58$\pm$0.05  &0.099  \\   
0235+164      &02/09/2008   &16.97$\pm$0.04 &17.93$\pm$0.04  &18.08$\pm $0.05  &18.2$\pm$0.05   &18.57$\pm$0.07  &18.91$\pm$0.05  &0.262  \\   
0347--211     &15/10/2008   &$>$19.0        &19.52$\pm$0.2   &19.75$\pm $0.34  &$>$20.15        &$>$19.93        &$>$20.66        &0.259  \\       
0426--380     &27/10/2008   &...            &...             &...              &...             &16.71$\pm$0.1   &...             &0.082   \\   
0454-234      &26/10/2008   &17.01$\pm$0.05 &17.48$\pm$0.03  &16.76$ \pm$0.03  &16.90$\pm$0.03  &16.98$\pm$0.05  &17.46$\pm$0.03  &0.157   \\   
0528+134      &02/10/2008   &...            &...             &....             &$> $21.05(1.8$\sigma$) &...      &...             &2.78   \\      
0820+560      &20/12/2008   &18.14$\pm$0.06 &18.34$\pm$0.04  &17.4$ \pm$0.03   &17.5$\pm$0.03   &17.6$\pm$0.03   &17.88$\pm$0.03  &0.209  \\   
0920+44       &18/01/2009   &17.29$\pm$0.05 &17.66$\pm$0.03  &17.04$ \pm$0.03  &18.32$\pm$0.06  &20.79$\pm$0.26  &19.98$\pm$0.11  &0.068 \\   
1454--354$_1$ &07/01/2008   &...            &...             &....             &...             &18.43$\pm$0.03  &...             &0.341  \\       
1454--354$_2$ &12/09/2008   &16.79$\pm$0.1  &17.62$\pm$0.06  &16.8$\pm $0.05   &17.15$\pm$0.06  &17.59$\pm$0.15  &18.12$\pm$0.07  &0.341  \\   
1502+106      &08/08/2008   &16.62$\pm$0.02 &17.08$\pm$0.02  &16.36$\pm $0.01  &16.63$\pm$0.02  &16.72$\pm$0.02  &16.9$\pm$0.01   &0.106  \\   
1520+319      &12/11/2008   &19.9$\pm$0.29  &20.18$\pm$0.18  &19.13$\pm $0.11  &19.7$\pm$0.13   &20.19$\pm$0.17  &21.18$\pm$0.21  &0.078  \\   
1633+382      &22/02/2007   &17.97$\pm$0.1  &18.01$\pm$0.04  &16.99$\pm $0.03  &17.47$\pm$0.04  &18.46$\pm$0.08  &19.31$\pm$0.09  &0.037  \\   
2023--07      &08/12/2008   &18.09$\pm$0.11 &18.57$\pm$0.07  &17.76$\pm $0.05  &17.73$\pm$0.04  &17.92$\pm$0.05  &18.16$\pm$0.04  &0.129  \\   
2227--088     &28/04/2005   &18.13$\pm$0.11 &18.51$\pm$0.12  &17.41$\pm $0.05  &17.67$\pm$0.05  &17.86$\pm$0.06  &18.86$\pm$0.06  &0.17  \\   
2251+158      &08/08/2008   &15.1$\pm$0.02  &15.71$\pm$0.01  &15.04$\pm $0.01  &15.28$\pm$0.01  &15.32$\pm$0.02  &15.57$\pm$0.01  &0.335 \\   
2325+093      &03/06/2008   &...            &...             &17.76$\pm$0.02   &...             &...             &...                        &0.204  \\       
\hline
\hline
\end{tabular}
\vskip 0.4 true cm
\caption{Results of the UVOT analysis.
The given magnitudes are not corrected for Galactic extinction.
The value of $A_V$ is taken from Schlegel et al. (1998)
}
\label{uvot}
\end{table*}
%

\begin{figure}
\vskip -0.2 cm
\psfig{figure=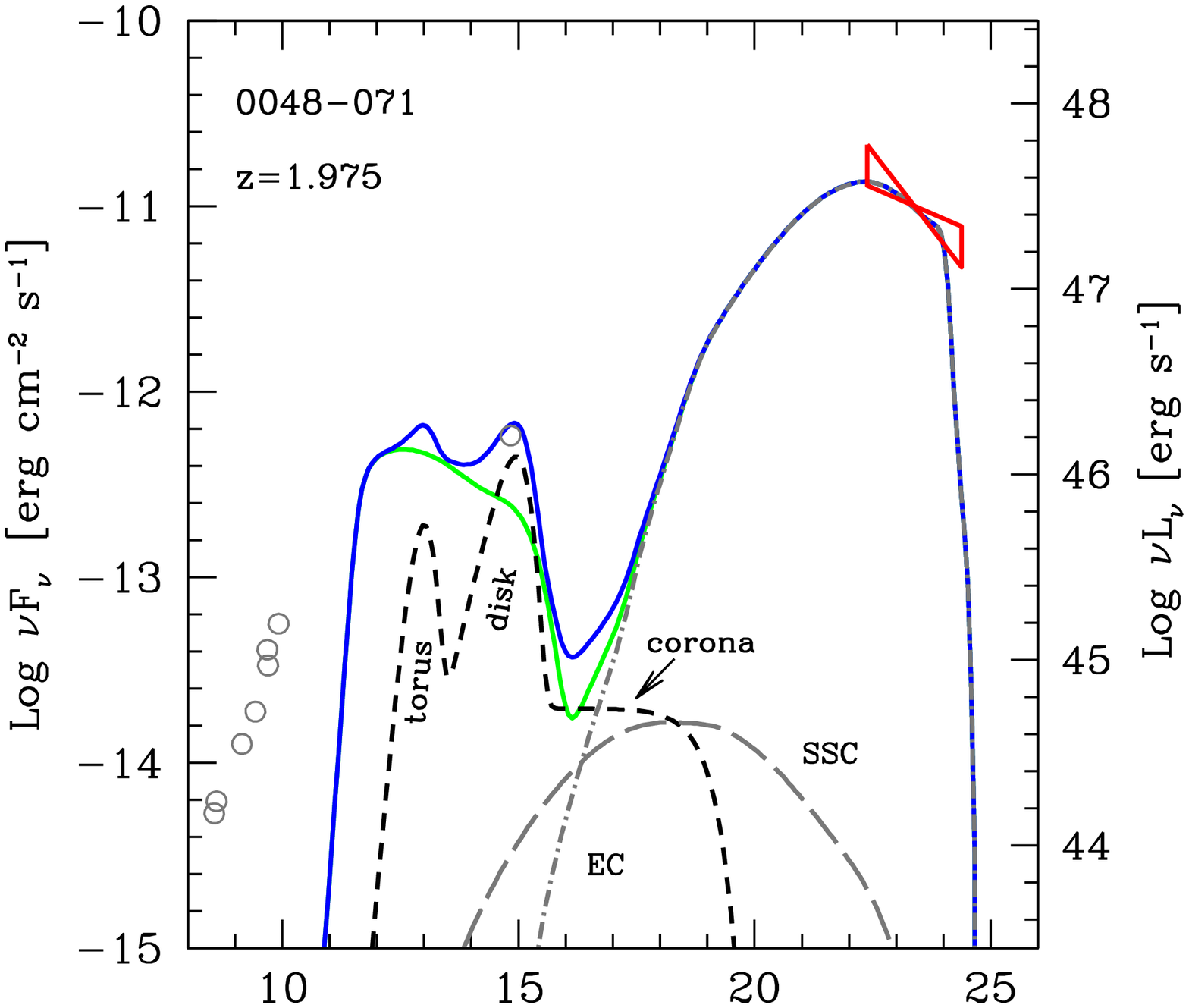,width=9cm,height=7cm}
\vskip -1.4 cm
\psfig{figure=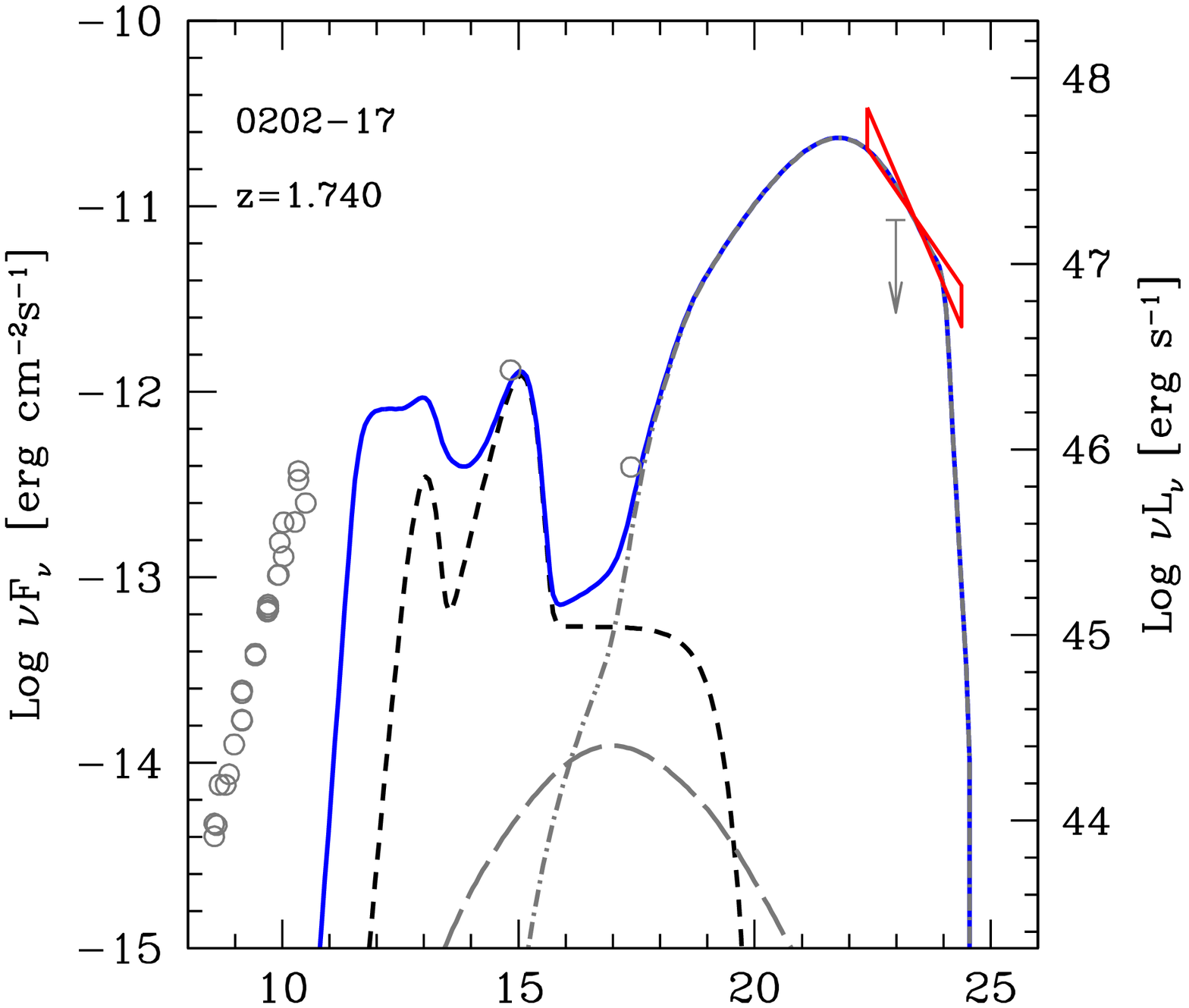,width=9cm,height=7cm}
\vskip -1.4 cm
\psfig{figure=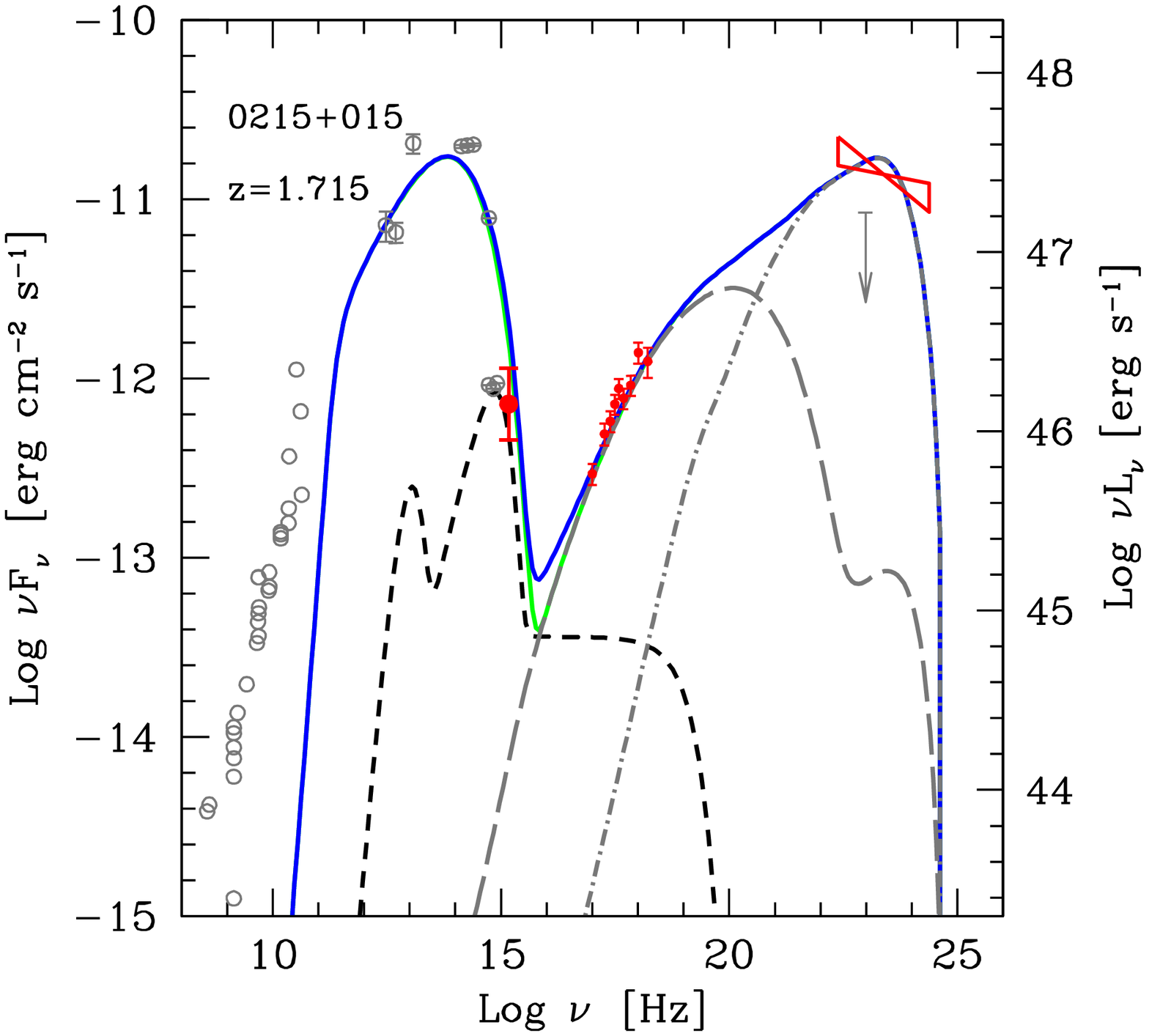,width=9cm,height=7cm}
\vskip -0.6 cm
\caption{SED of PKS 0048--071, 0202--17 and 
PKS 0215+015, together with the fitting models,
with parameters listed in Tab. \ref{para}.
{\it Fermi} and {\it Swift} data are indicated
by dark grey symbols (red in the electronic version),
while archival data (from NED) are in light grey.
The short--dashed line is the emission from the IR torus, the accretion disk
and its X--ray corona; the long--dashed line is the SSC contribution and
the dot--dashed line is the EC emission.
The solid light grey line (green in the electronic version) is the
non thermal flux produced by the jet, the solid dark grey line (blue
in the electronic version) is the sum of the non--thermal and thermal 
components.
}
\label{f1}
\end{figure}

\begin{figure}
\vskip -0.2cm
\psfig{figure=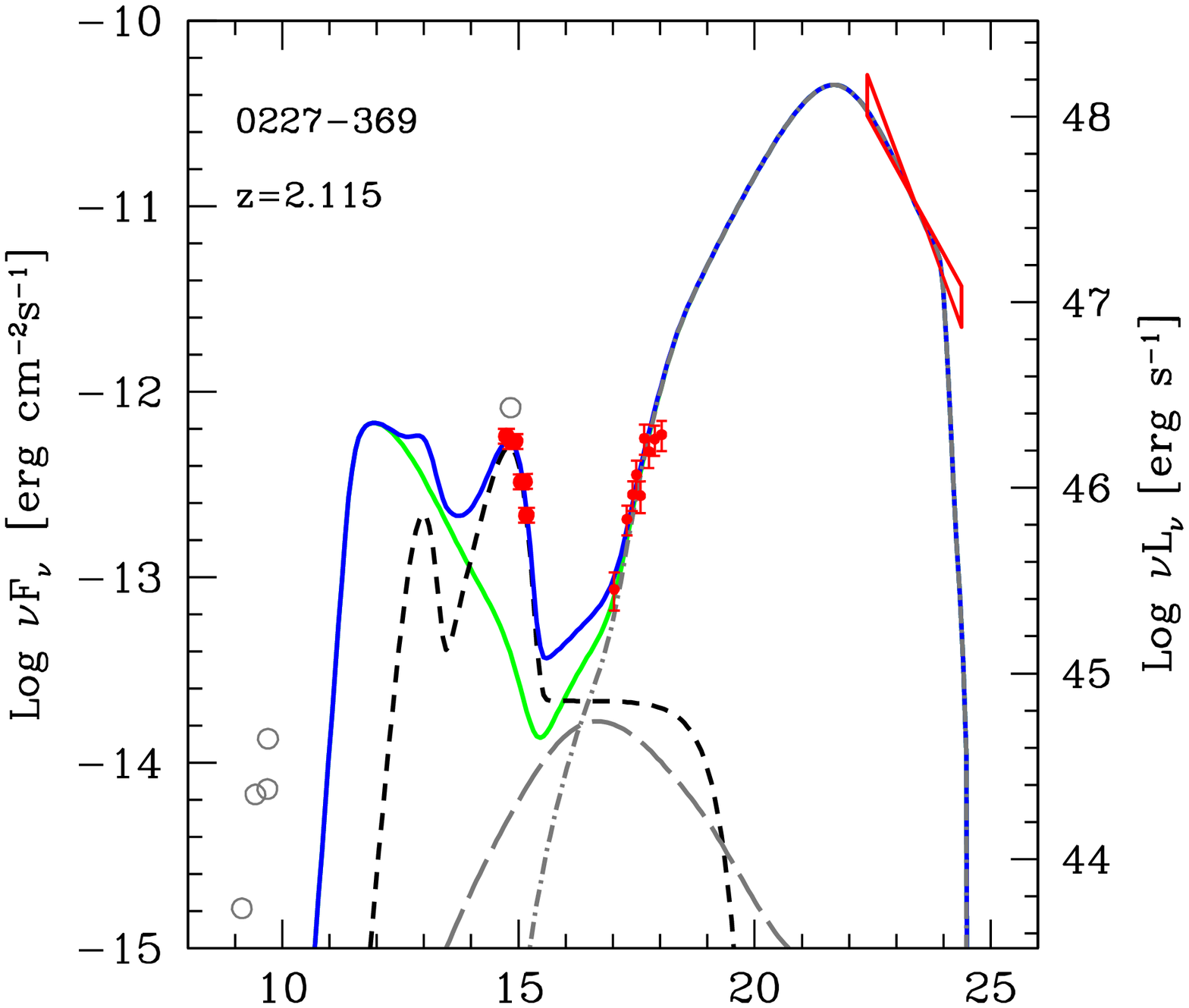,width=9cm,height=7cm}
\vskip -1.4 cm
\psfig{figure=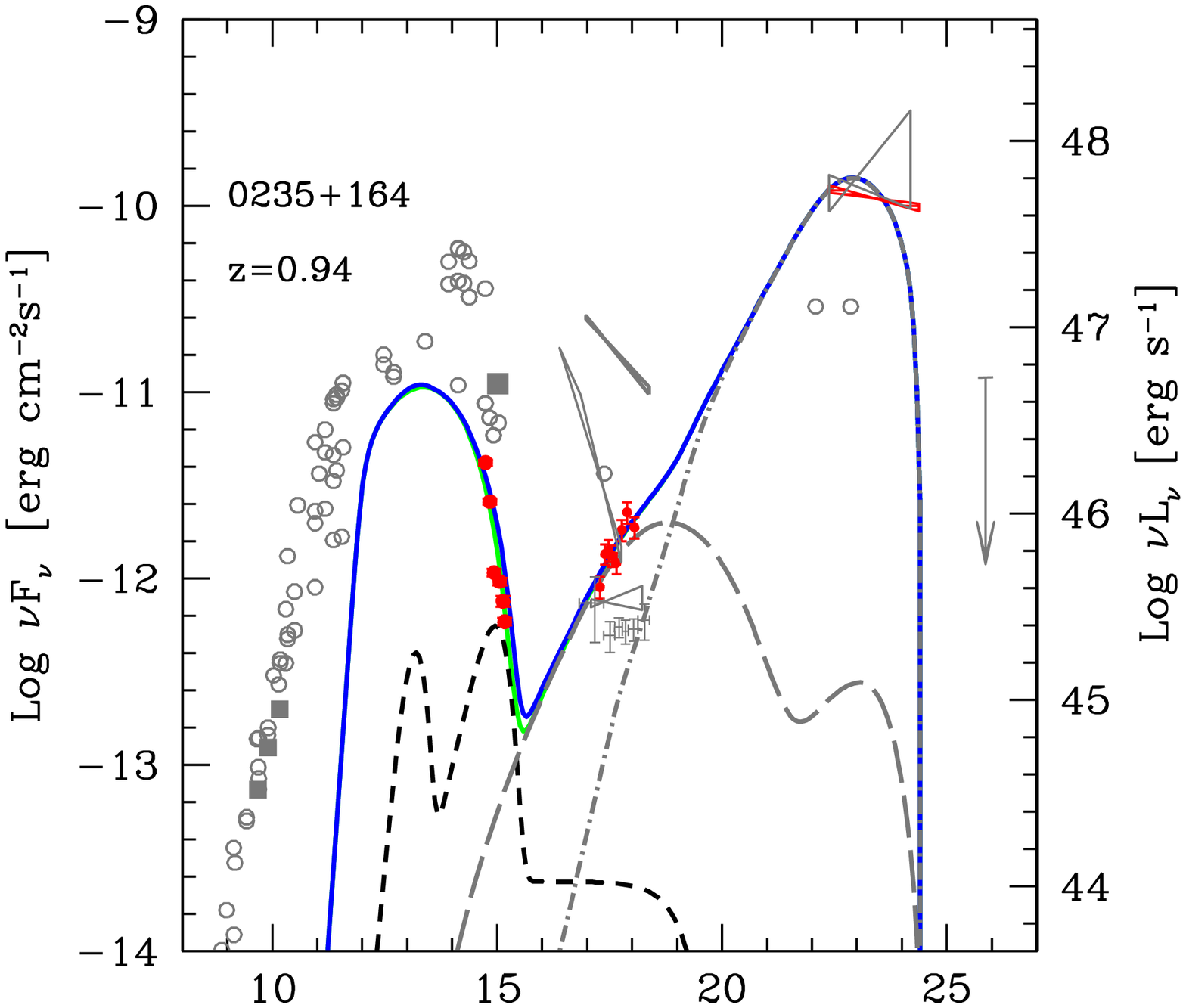,width=9cm,height=7cm}
\vskip -1.4cm
\psfig{figure=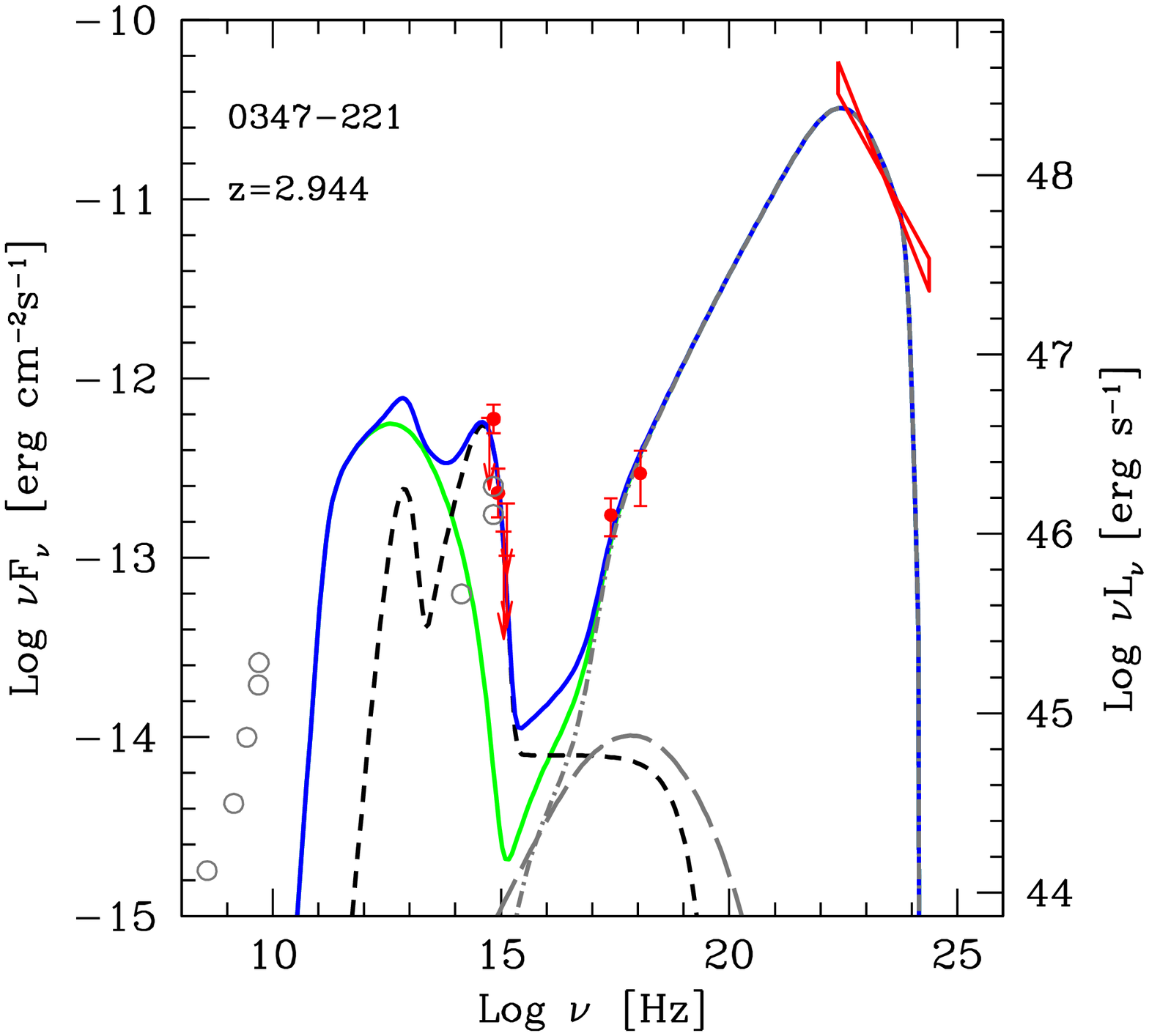,width=9cm,height=7cm}
\vskip -0.6 cm
\caption{SED of PKS 0227--369; A0 0235+164 and  PKS 0347--221.
Symbols and lines as in Fig. \ref{f1}.
}
\label{f2}
\end{figure}

\begin{figure}
\vskip -0.2cm
\psfig{figure=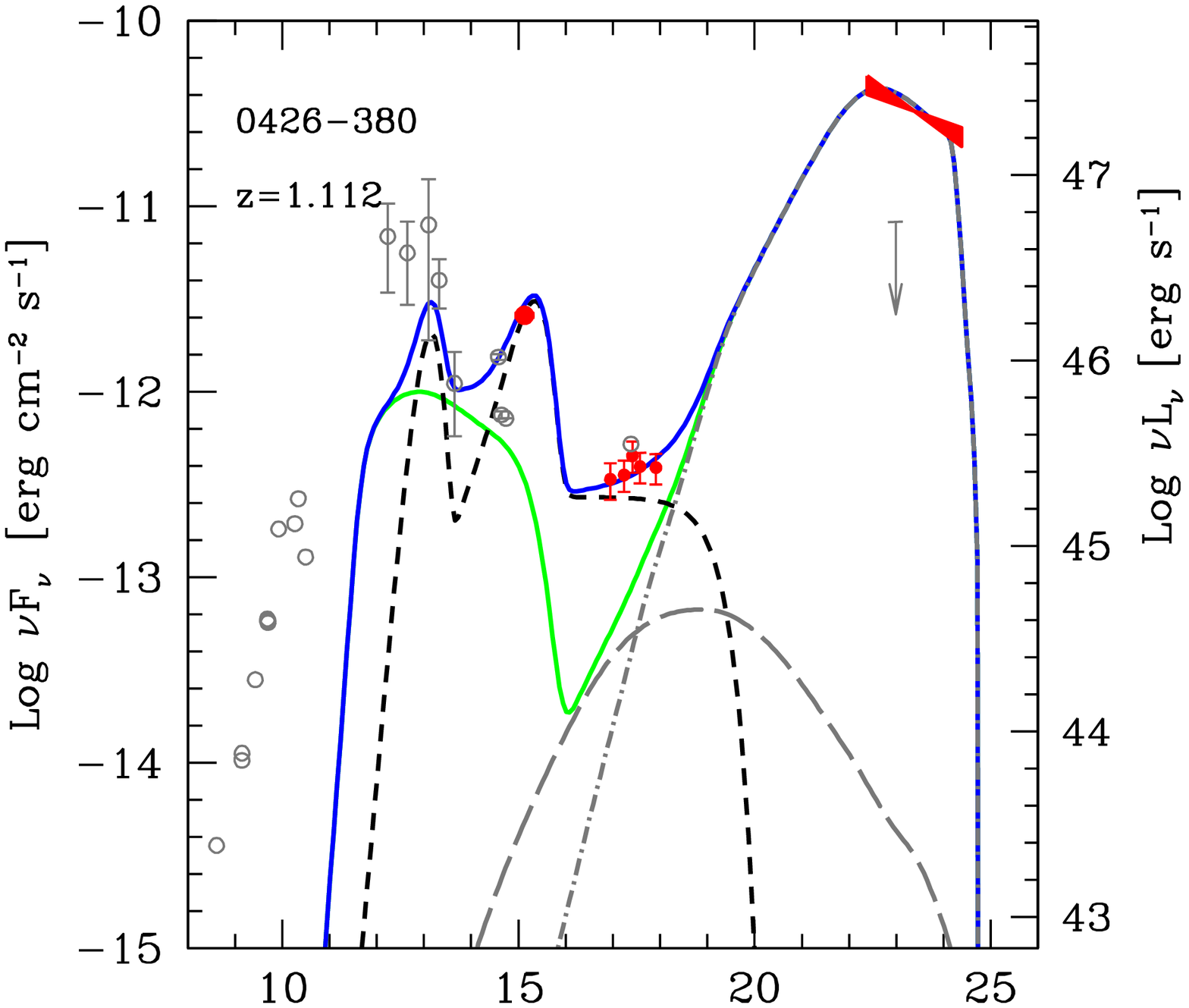,width=9cm,height=7cm}
\vskip -1.4cm
\psfig{figure=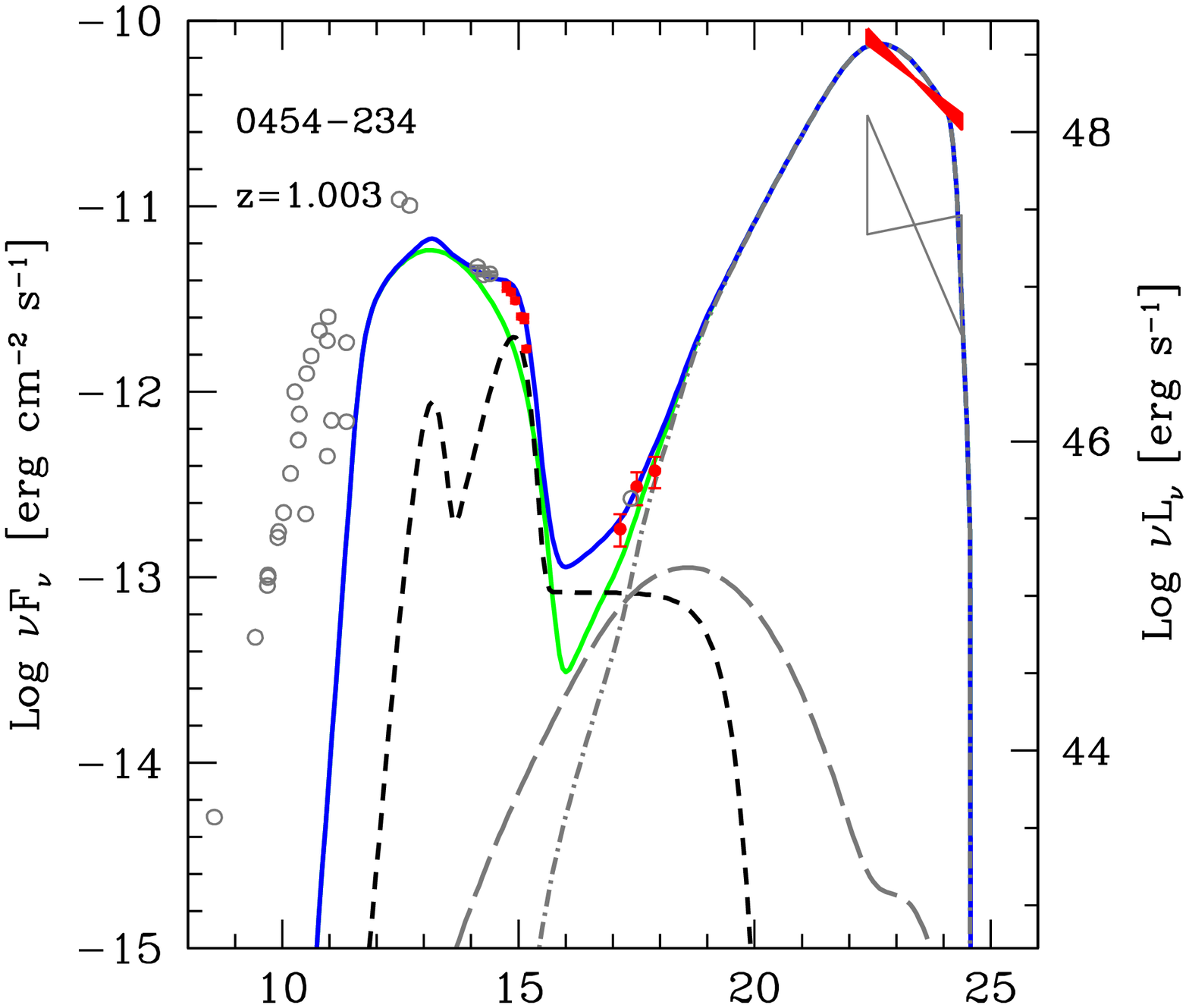,width=9cm,height=7cm}
 \vskip -1.4cm
\psfig{figure=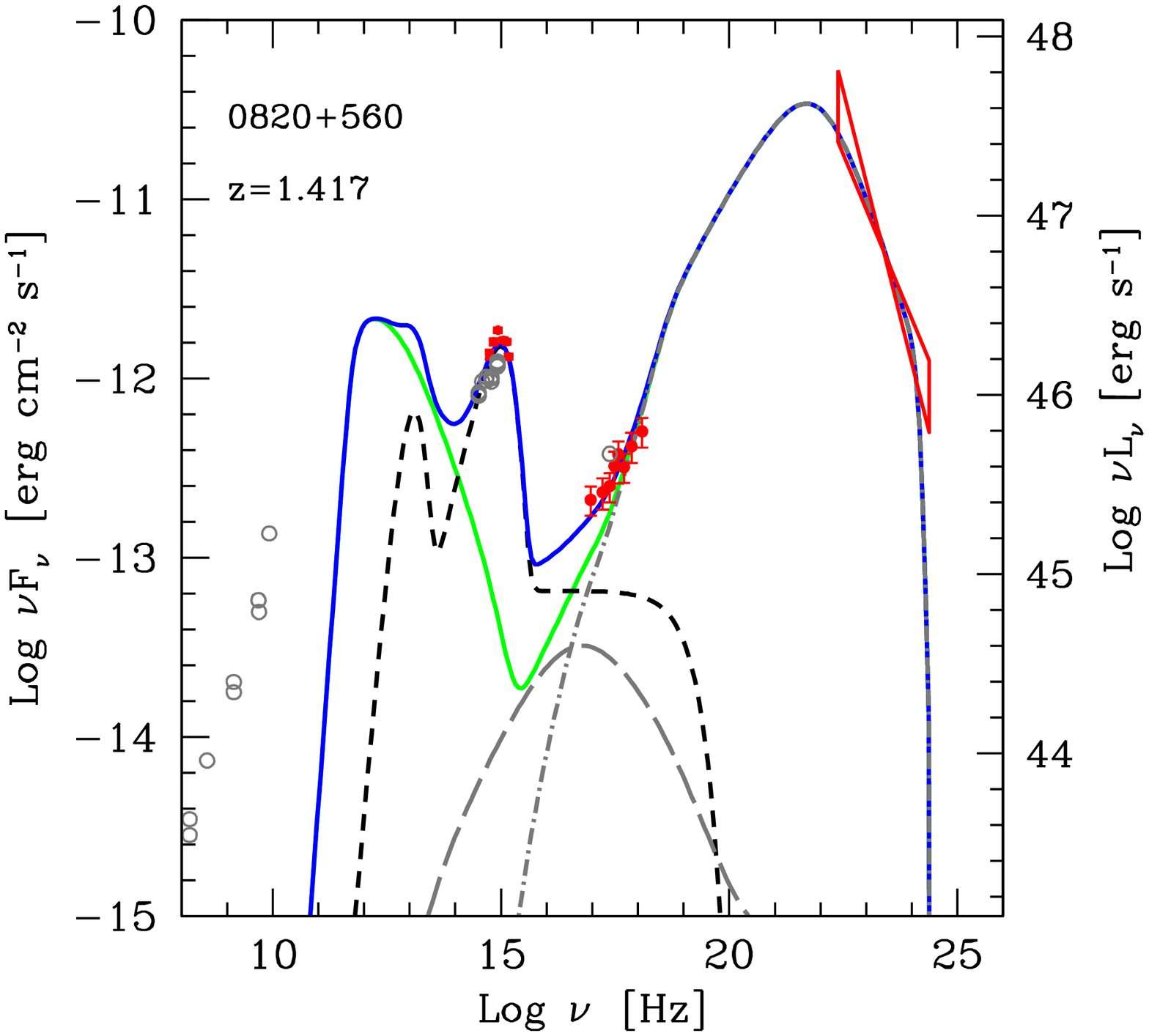,width=9cm,height=7cm}
\vskip -0.6 cm
\caption{SED of PKS 0426--380, PKS 0445--234 and PKS 0820+560
Symbols and lines as in Fig. \ref{f1}.
}
\label{f3}
\end{figure}

\begin{figure}
\vskip -0.2cm
\psfig{figure=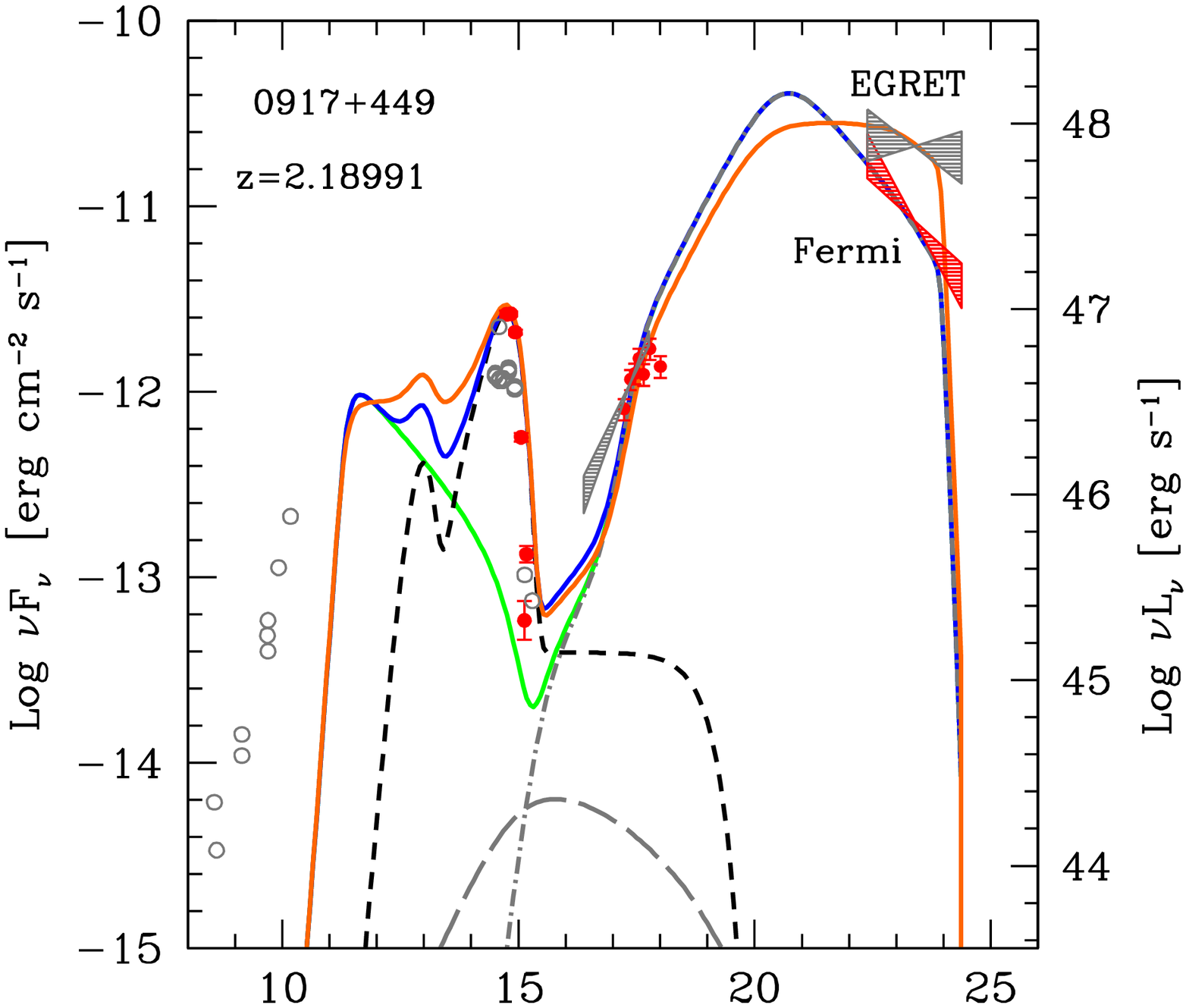,width=9cm,height=7cm}
\vskip -1.4cm
\psfig{figure=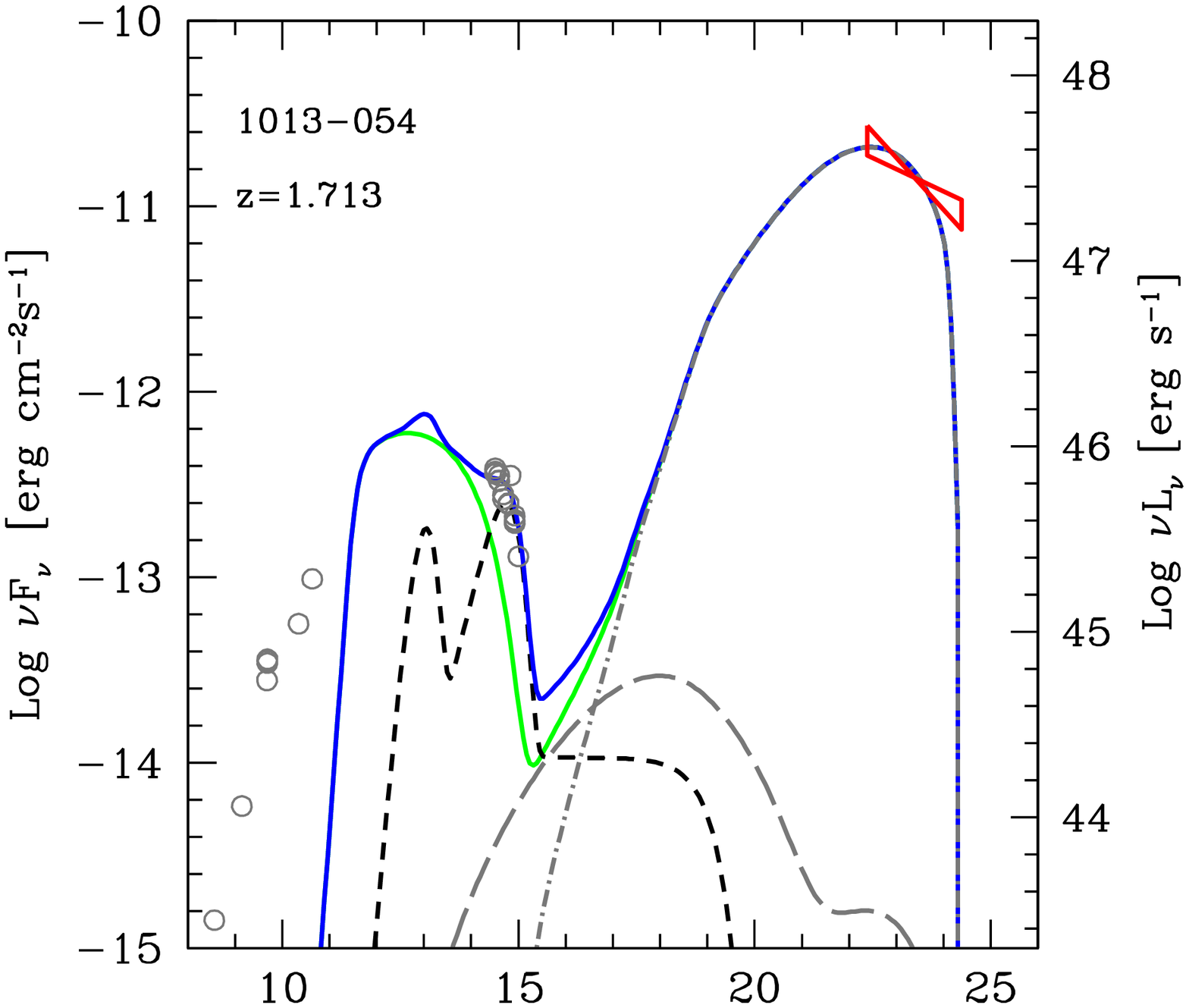,width=9cm,height=7cm}
\vskip -1.4cm
\psfig{figure=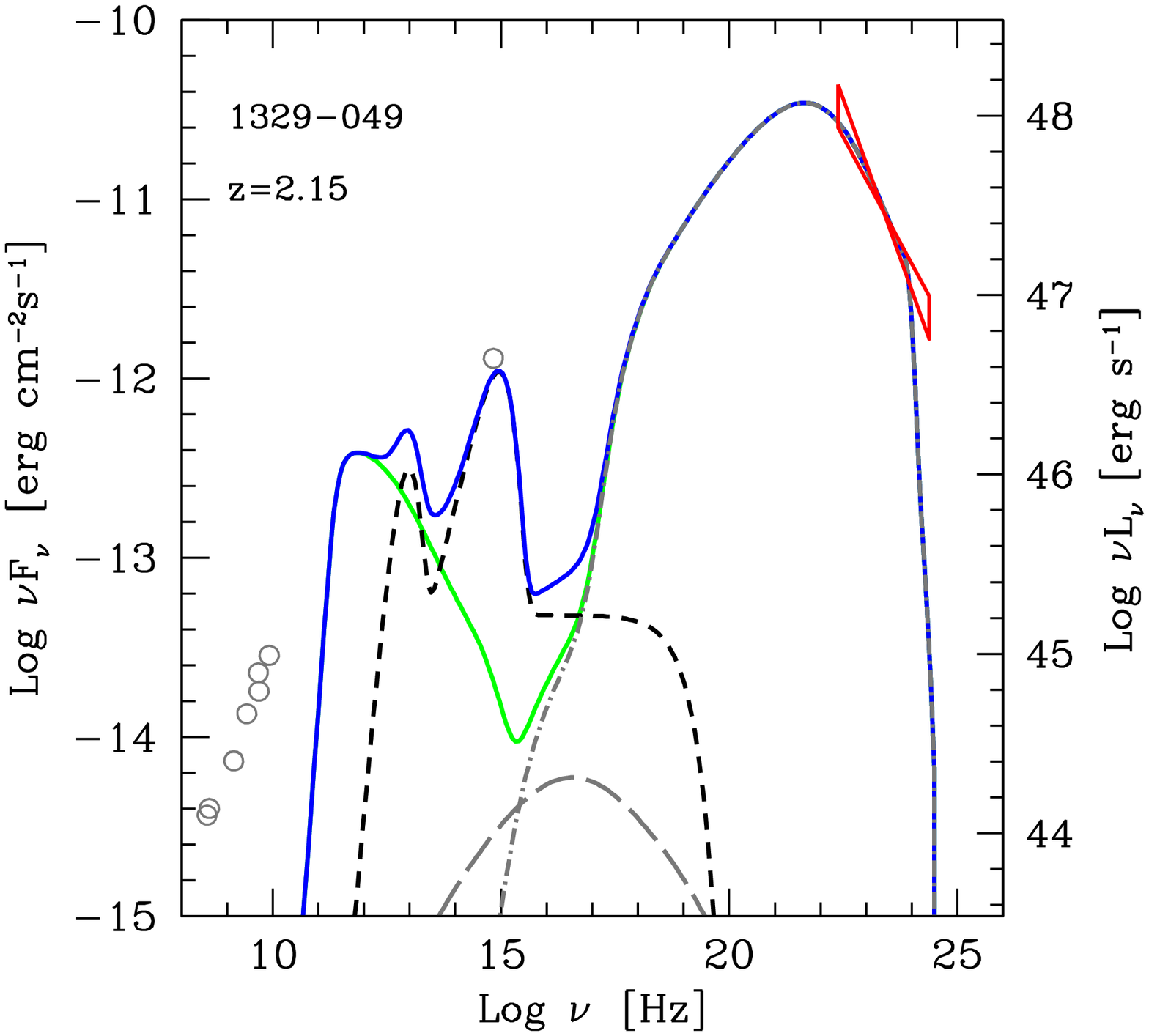,width=9cm,height=7cm}
\vskip -0.6 cm
\caption{SED of 0917+449=S4 0920+44, 1013--054 and 1329-049. 
Symbols and lines as in Fig. \ref{f1}.
For PKS 0917+449 we show two models, corresponding to 
the {\it Fermi/Swift} data and to the older EGRET data.
}
\label{f3bis}
\end{figure}

\begin{figure}
\vskip -0.2cm
\psfig{figure=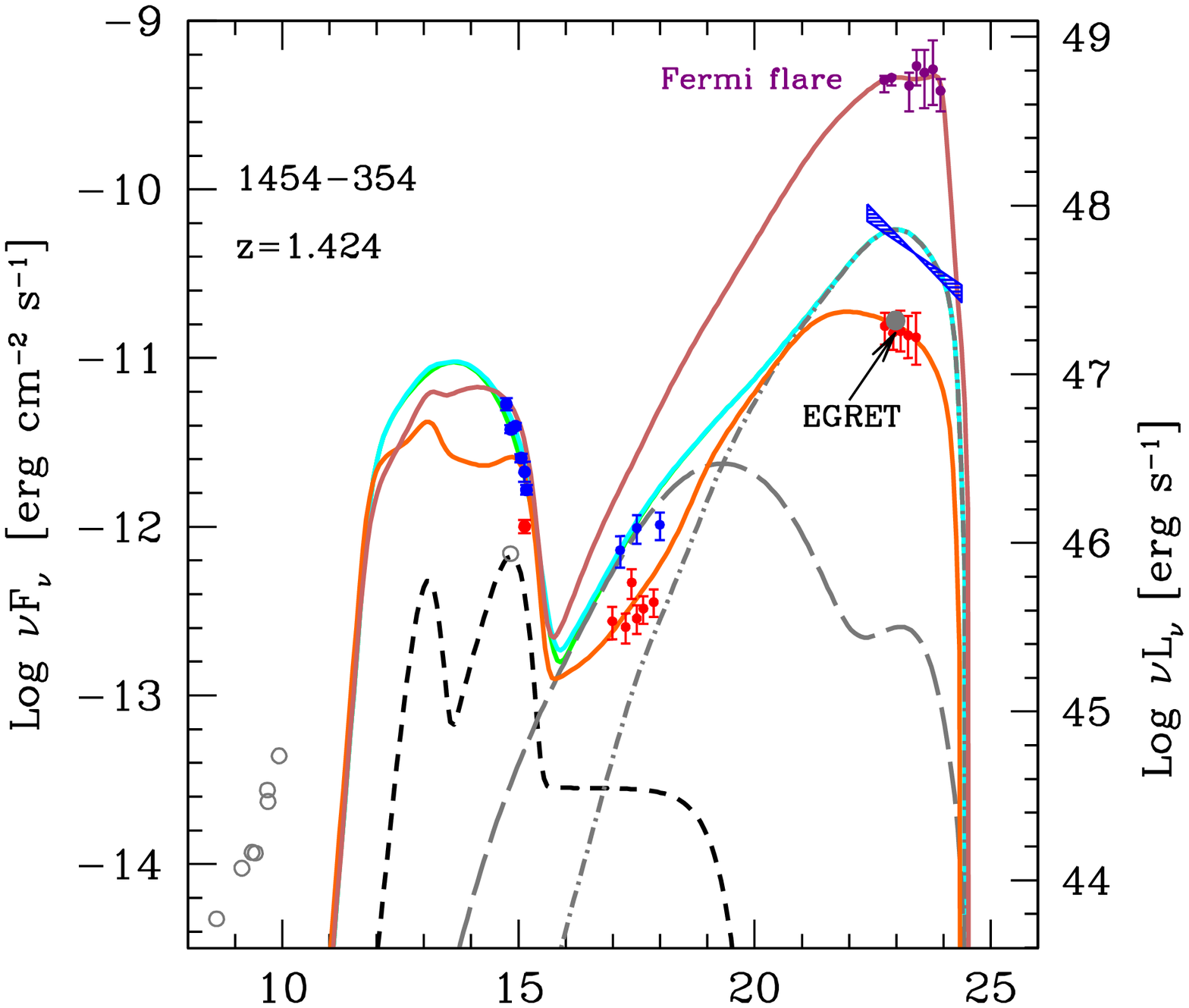,width=9cm,height=7cm}
\vskip -1.4cm
\psfig{figure=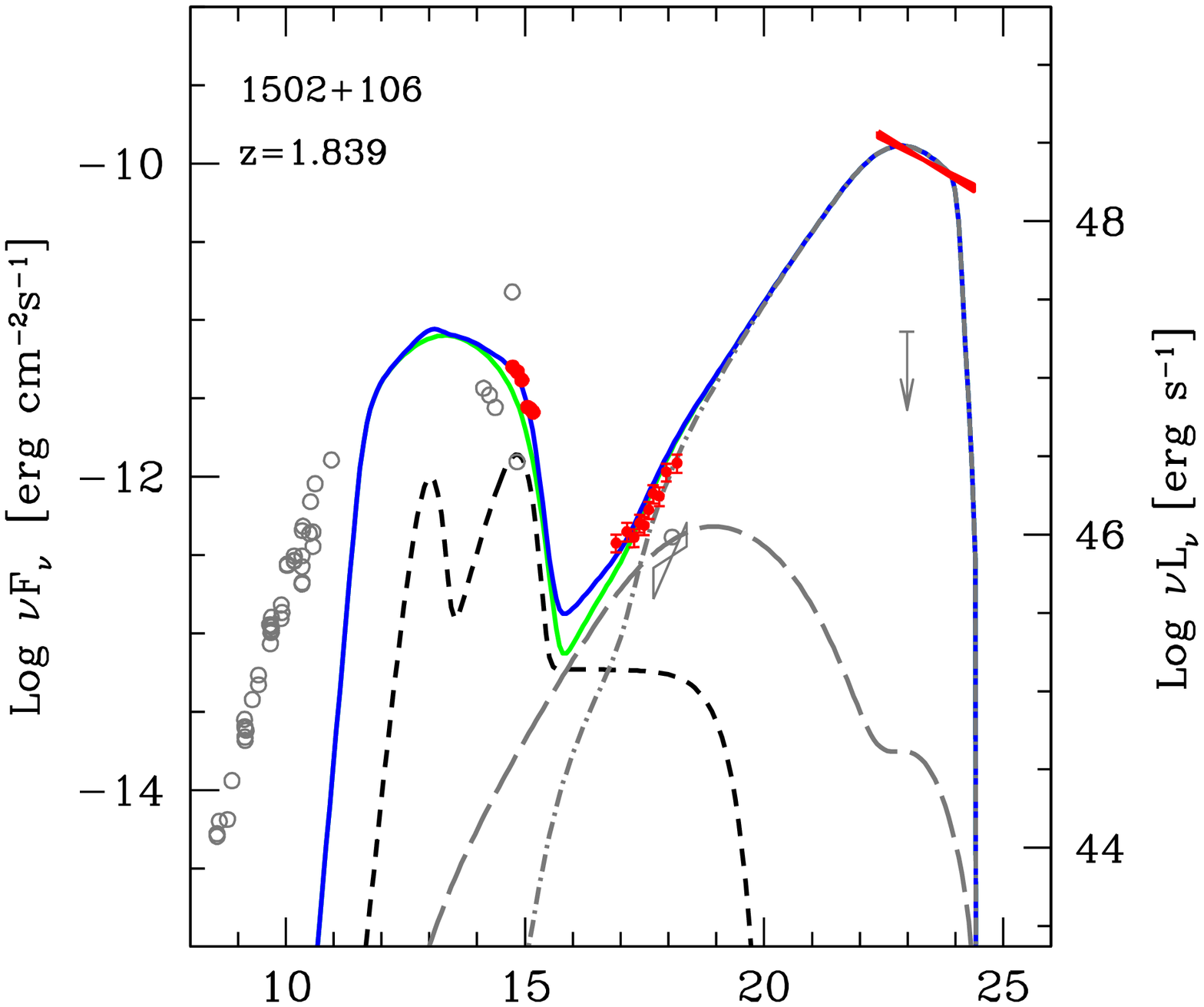,width=9cm,height=7cm}
\vskip -1.4 cm
\psfig{figure=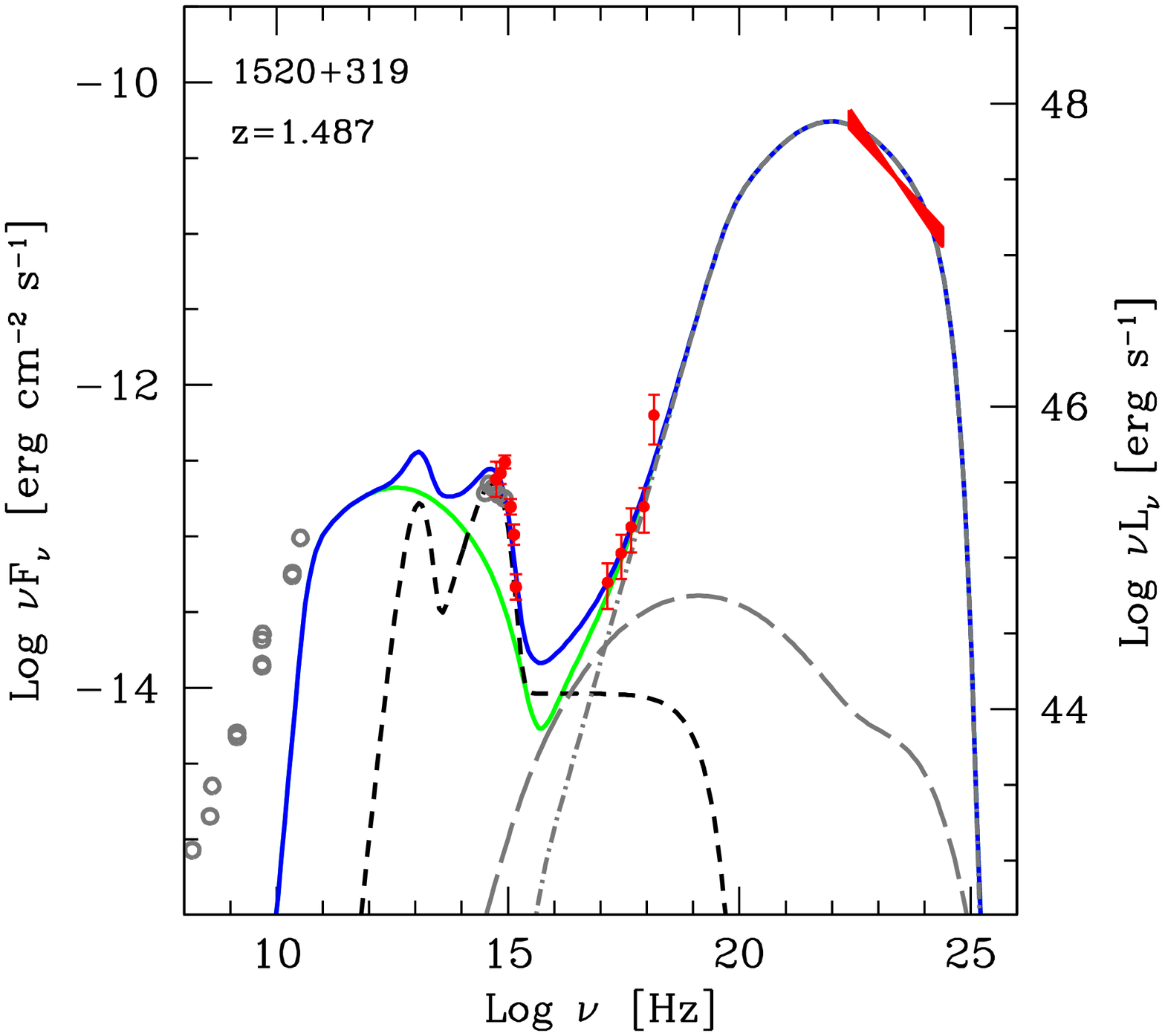,width=9cm,height=7cm}
\vskip -0.5 cm
\caption{
SED of 1424--354, PKS 1502+106; PKS 1520+319
Symbols and lines as in Fig. \ref{f1}.
}
\label{f4}
\end{figure}

\begin{figure}
\vskip -0.2cm
\psfig{figure=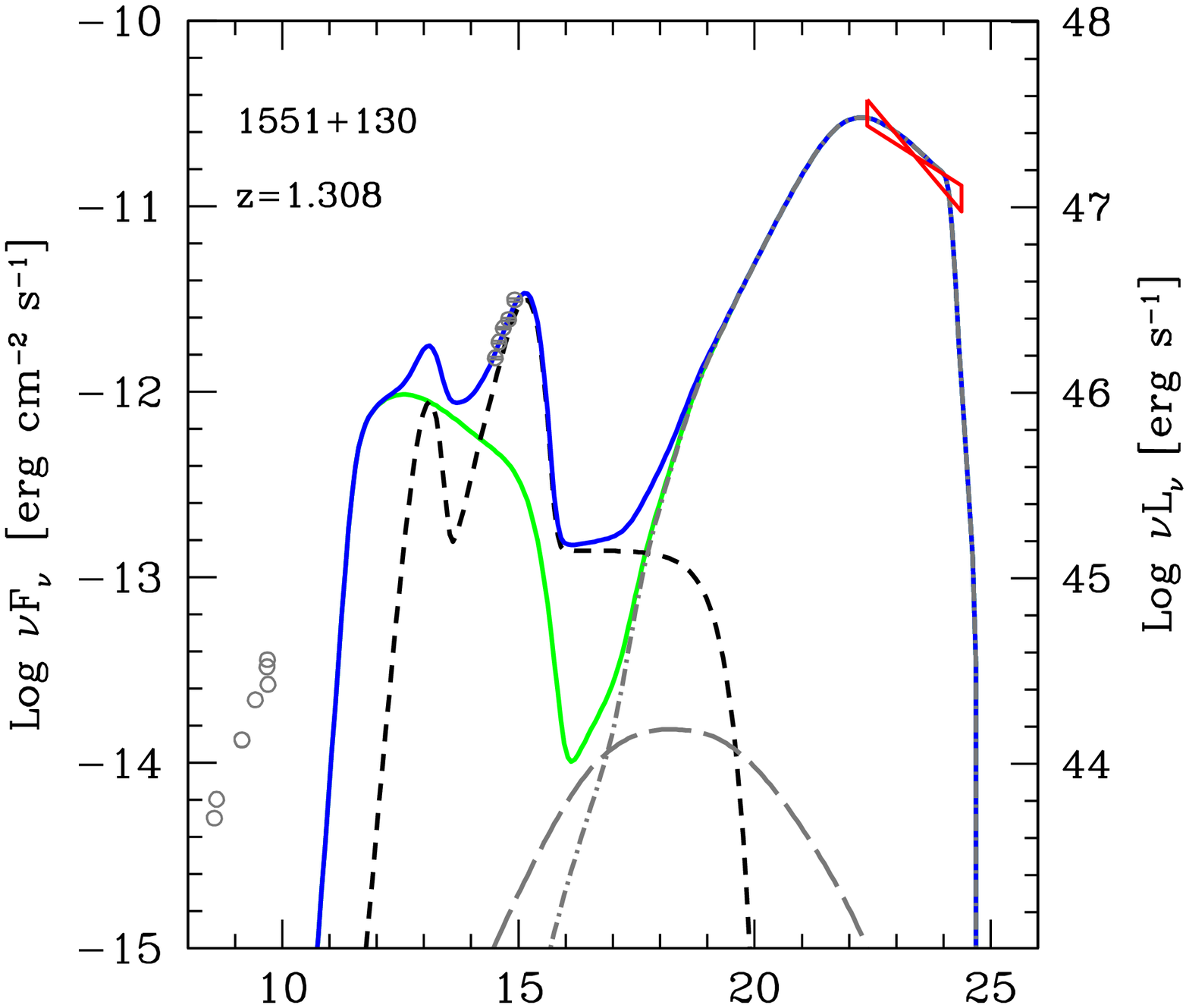,width=9cm,height=7cm}
\vskip -1.4 cm
\psfig{figure=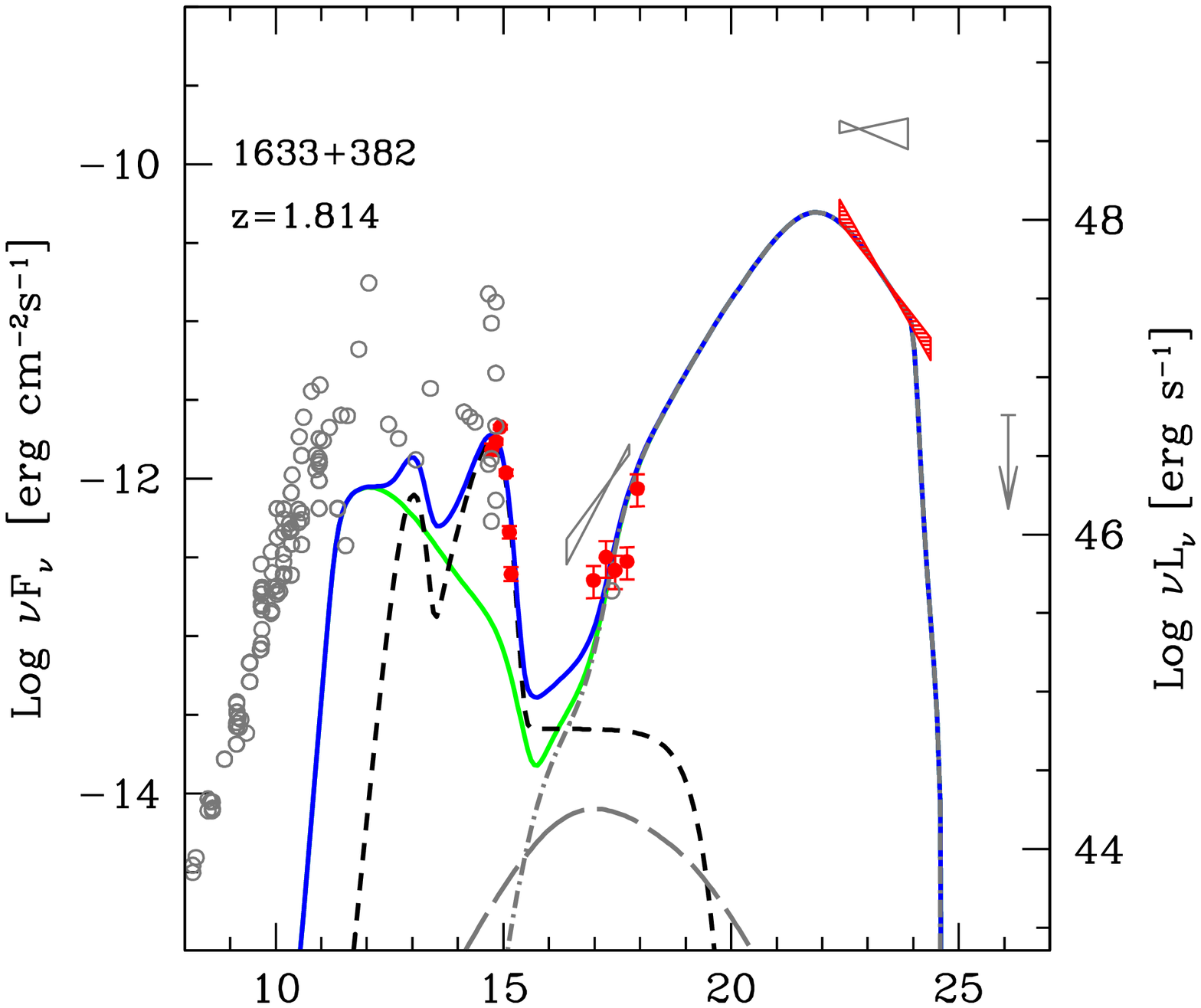,width=9cm,height=7cm}
\vskip -1.4 cm
\psfig{figure=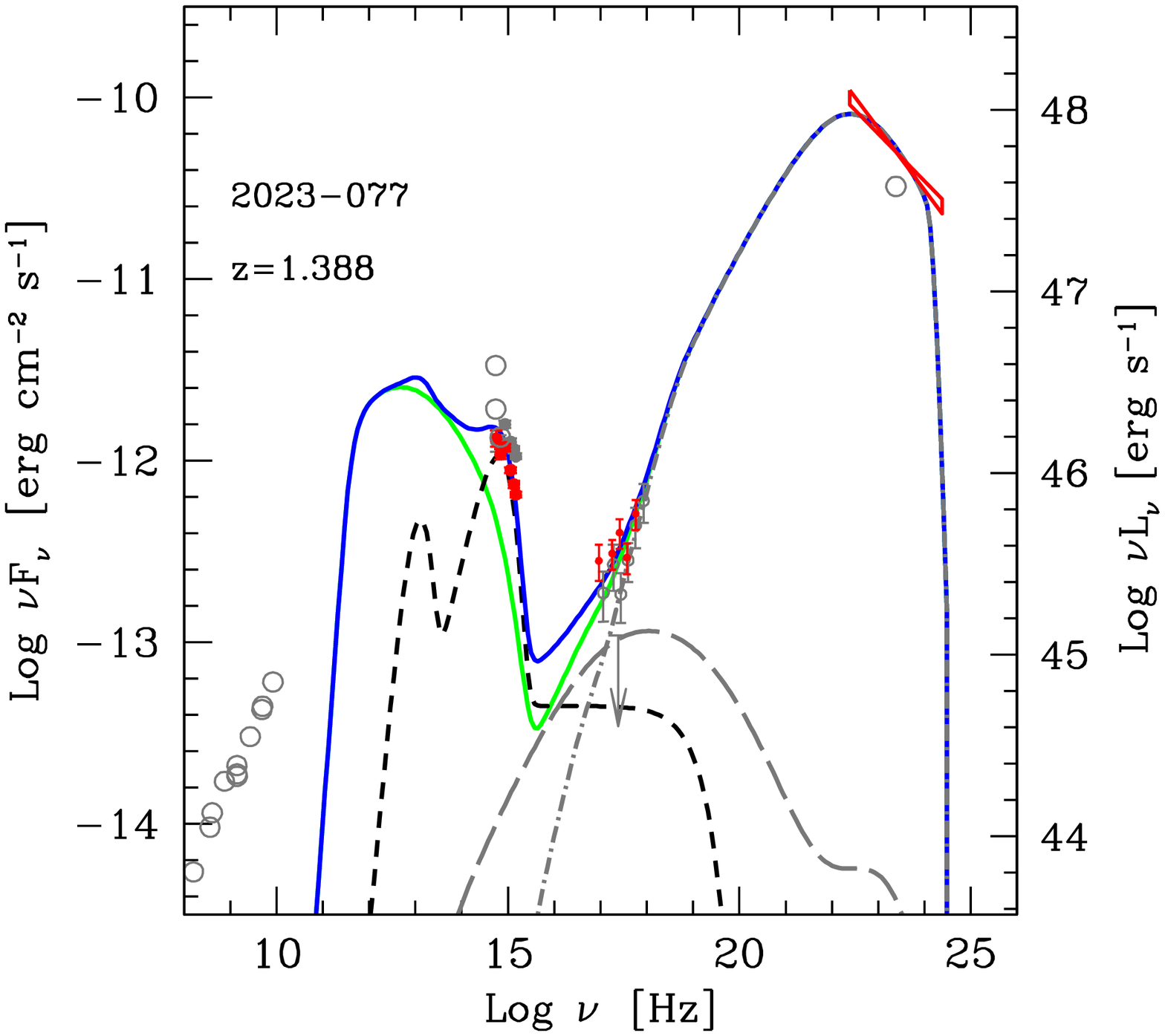,width=9cm,height=7cm}
\vskip -0.5 cm
\caption{SED of PKS 1551+130,
PKS 1633+382 and PKS 2023--077.
Symbols and lines as in Fig. \ref{f1}.
}
\label{f5}
\end{figure}

\begin{figure}
\vskip -0.2cm
\psfig{figure=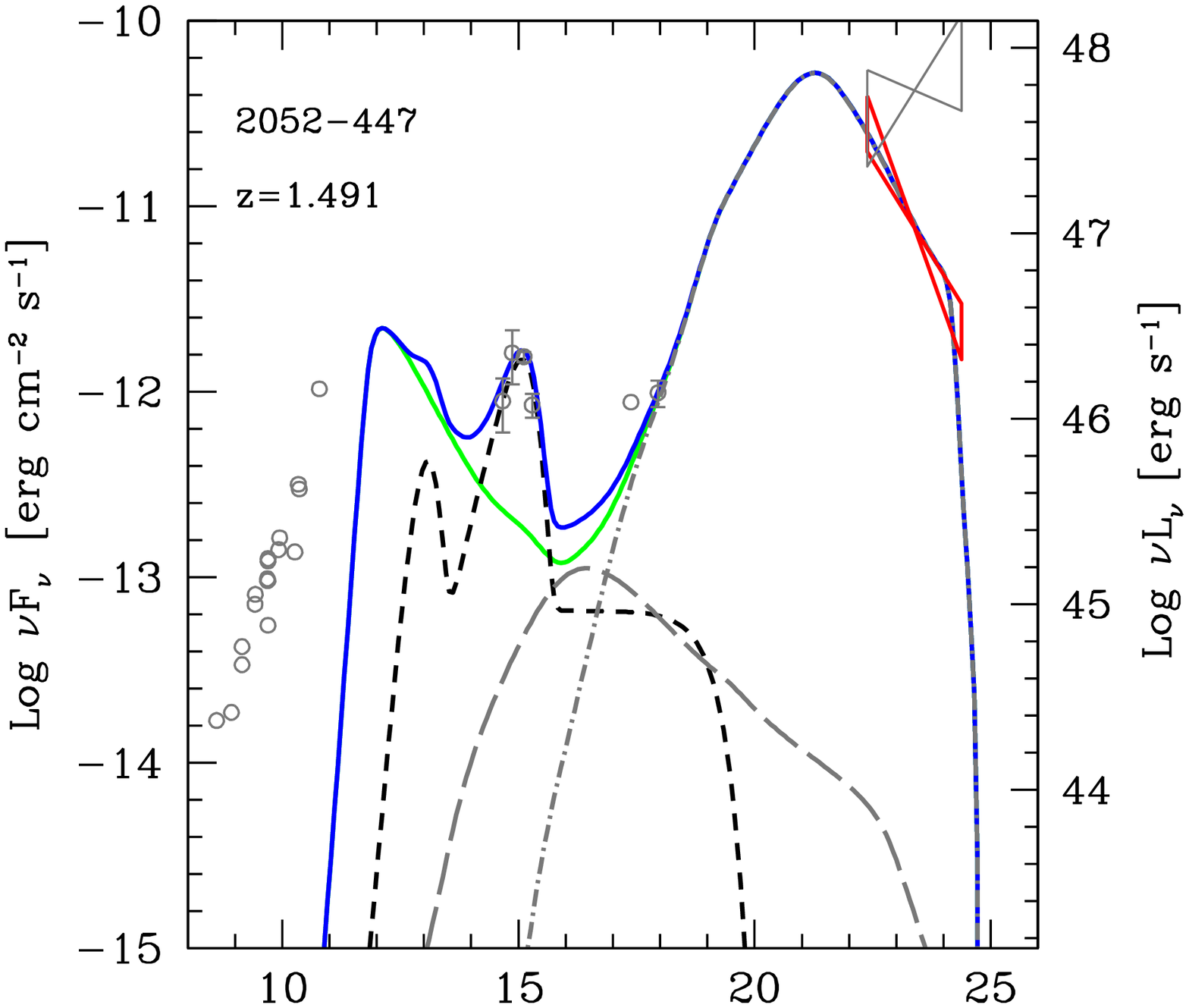,width=9cm,height=7cm}
\vskip -1.4 cm
\psfig{figure=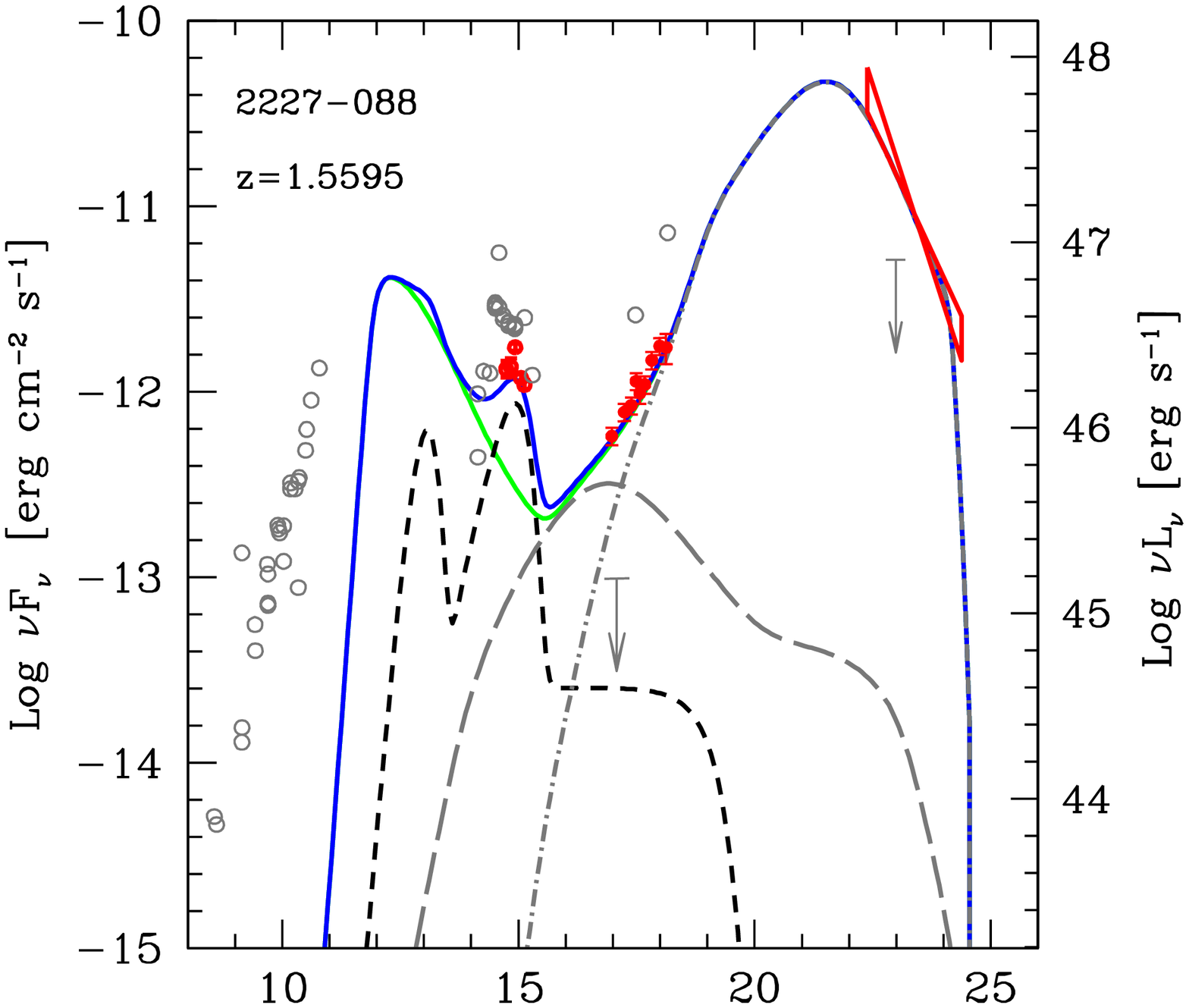,width=9cm,height=7cm}
\vskip -1.4 cm
\psfig{figure=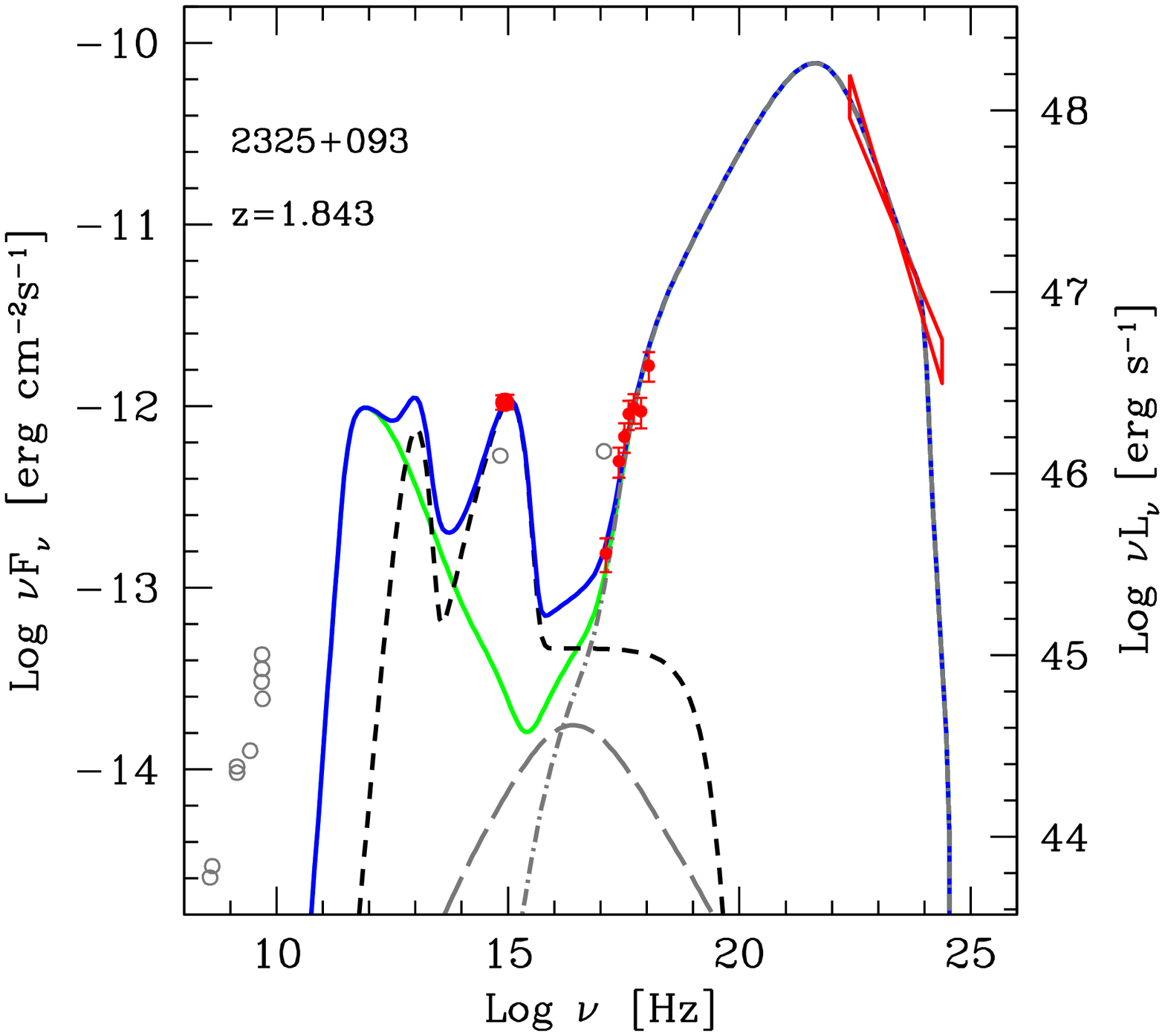,width=9cm,height=7cm}
\vskip -0.5 cm
\caption{SED of PKS 2052--47, PKS 2227--088, PKS 2052--447
and PKS 2325+093.
Symbols and lines as in Fig. \ref{f1}.
}
\label{f6}  
\end{figure}

\begin{figure*}
\vskip -1cm
\psfig{figure=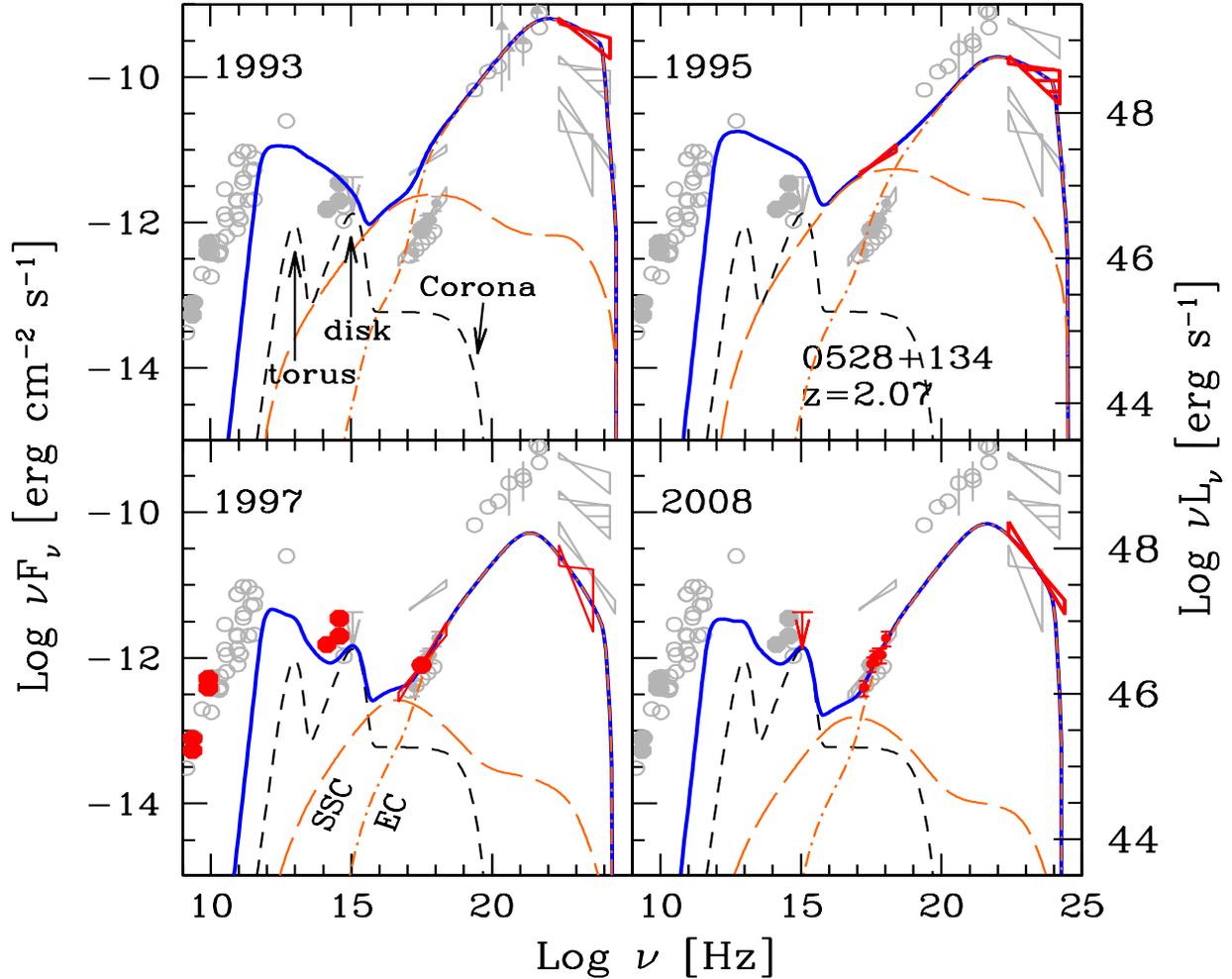,width=18cm,height=16cm}
\vskip -1.5 cm
\caption{SED of PKS 0528+134 at four epochs.
For the first three, the $\gamma$--ray data comes from EGRET.
For the 2008 SED, the $\gamma$--ray data are from {\it Fermi}.
Symbols and lines as in Fig. \ref{f1}.
}
\label{f0528}
\end{figure*}

\begin{figure*}
\vskip -1cm
\psfig{figure=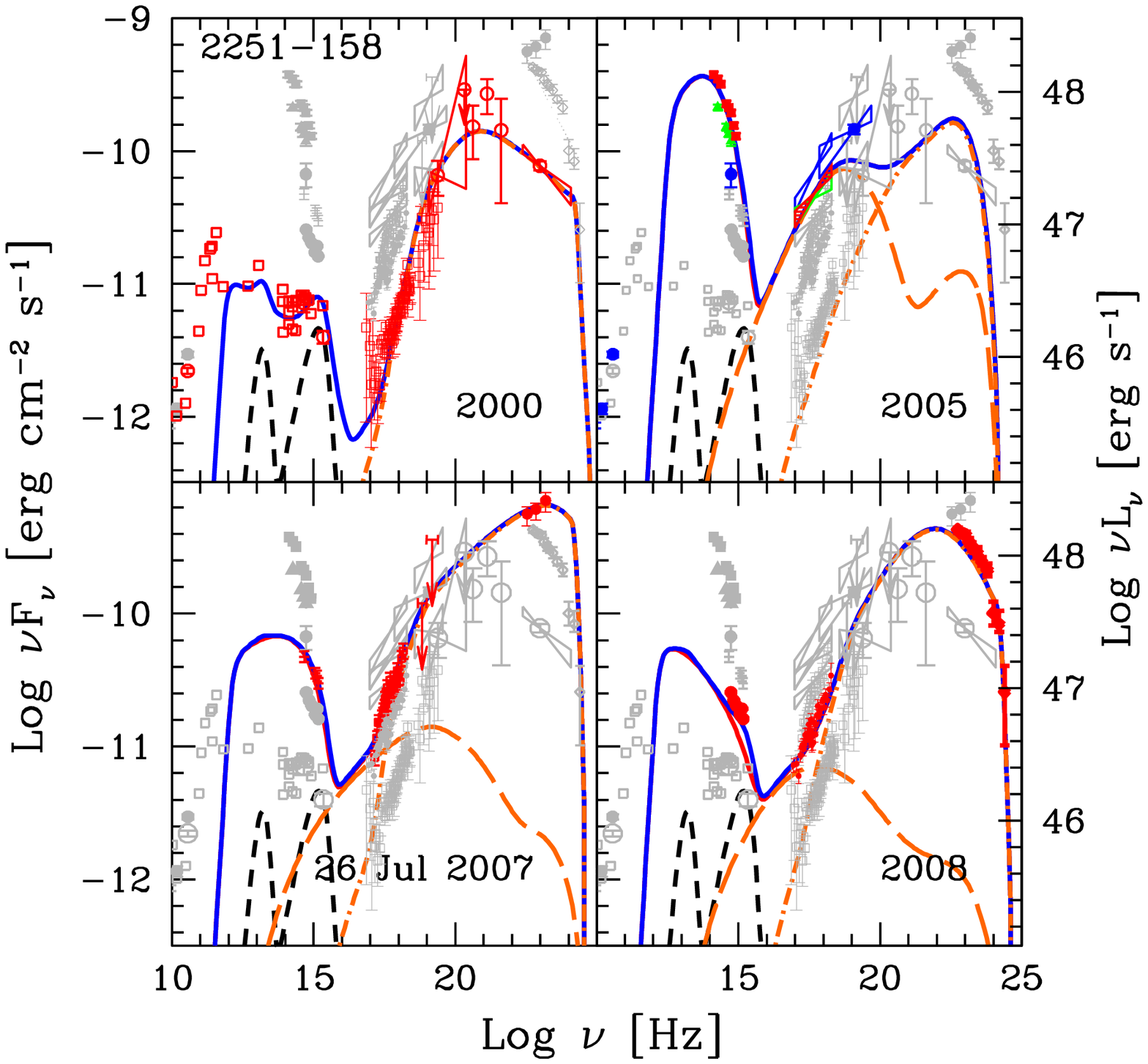,width=18cm,height=16cm}
\vskip -1.5 cm
\caption{SED of PKS 2251--158=3C 454.3 at four epochs.
For the July 2007 SED, $\gamma$--ray data comes from {\it AGILE}
(Vercellone et al. 2009).
For the 2008 SED, the $\gamma$--ray data are from {\it Fermi}.
Symbols and lines as in Fig. \ref{f1}.
}
\label{f7}  
\end{figure*}

\section{The model}

To interpret the overall SED of our sources we use a relatively simple,
one zone, homogeneous synchrotron and Inverse Compton model, described
in detail in GT09.
This model aims at accounting in a simple yet accurate way the
several contributions to the radiation energy density
produced externally to the jet, and their dependence upon the
distance of the emitting blob to the black hole.
Here we summarise the main characteristics of the model.
The emitting region, of size $r_{\rm diss}$ and moving with a bulk Lorentz factor 
$\Gamma$, is located at a distance $R_{\rm diss}$ from the 
black hole of mass $M$.
The accretion disk emits a bolometric luminosity $L_{\rm d}$.
The jet is assumed to be accelerating in its inner parts,
with $\Gamma\propto R^{1/2}$ ($R$ is the distance from the black hole),
up to a value $\Gamma_{\rm max}$. 
In the acceleration region the jet is assumed to be parabolic 
(following, e.g. Vlahakis \& K\"onigl 2004); beyond this point the jet
is conical with a semi--aperture angle $\psi$ (assumed to be 0.1 for all sources).

The energy particle distribution $N(\gamma)$ [cm$^{-3}$]
is calculated solving the continuity 
equation where particle injection, radiative cooling and pair production
(via the $\gamma$--$\gamma \to e^\pm$ process) are taken into account.
The injection function $Q(\gamma)$ [cm$^{-3}$ s$^{-1}$]
is assumed to be a smoothly joining broken power--law,
with a slope $Q(\gamma)\propto \gamma^{-{s_1}}$ and
$\gamma^{-{s_2}}$ below and above a break energy $\gamma_{\rm b}$:
\begin{equation}
Q(\gamma)  \, = \, Q_0\, { (\gamma/\gamma_{\rm b})^{-s_1} \over 1+
(\gamma/\gamma_{\rm b})^{-s_1+s_2} }
\label{qgamma}
\end{equation}

The total power injected into the source in the form of relativistic
electrons is $P^\prime_{\rm i}=m_{\rm e}c^2 V\int Q(\gamma)\gamma d\gamma$,
where $V=(4\pi/3)r_{\rm diss}^3$ is the volume of the emitting region,
assumed spherical.

We assume that the injection process lasts for a light crossing time
$r_{\rm diss}/c$, and calculate $N(\gamma)$ at this time.
The rationale for this choice is that even if injection could last longer,
adiabatic losses caused by the expansion of the source (that
is travelling while emitting) and the corresponding decrease 
of the magnetic field would make the observed flux to decrease. 
Therefore our calculated spectra correspond to the maximum of
a flaring episode.

Besides its own synchrotron radiation, the jet can scatter photons
produced externally to the jet by the accretion disk itself
(e.g. Dermer \& Schlickeiser 1993),
the broad line region (BLR; 
e.g. Sikora, Begelman \& Rees 1994) and a dusty torus
(see  B{\l}azejowski et al. 2000; Sikora et al. 2002).
We assume that the BLR (assumed for simplicity to be a thin spherical shell) 
is located at a distance $R_{\rm BLR}=10^{17}L_{\rm d, 45}^{1/2}$ cm,
emitting, in broad lines, a fraction $f_{\rm BLR}=0.1$ of the disk luminosity.
The square root dependence of $R_{\rm BLR}$ with $L_{\rm d}$ implies that the
radiation energy density of the broad line emission within the BLR is constant,
but is seen amplified by a factor $\sim\Gamma^2$ by the moving blob, as long
as $R_{\rm diss}<R_{\rm BLR}$.
A dusty torus, located at a distance 
$R_{\rm IR}=2.5\times 10^{18}L_{\rm d}^{1/2}$ cm,
reprocesses a fraction $f_{\rm IR}$ (of the order of 0.1--0.3) of $L_{\rm d}$
through dust emission in the far IR.
Above and below the accretion disk, in its inner parts, there is an X--ray emitting
corona of luminosity $L_{\rm X}$ (we fix it at a level of 30\% of $L_{\rm d}$).
Its spectrum is a power--law of energy index $\alpha_X=1$ ending with a exponential cut
at $E_{\rm c}=$150 keV.
We calculate the specific energy density (i.e. as a function of frequency) of 
all these external components as seen in the comoving frame, to properly calculate
the resulting Inverse Compton spectrum.

\section{Fitting guidelines}

In this section we illustrate some
considerations that are used as a guide for
the choice of the parameters.

\begin{itemize}

\item 
All our sources do have broad emission lines 
(including the two sources classified as BL Lacs, as discussed above).
The presence of visible broad lines ensures that the
ionising continuum, that we associate to the accretion disk, 
cannot be much lower that the observed optical--UV flux.

\item
The shape of the $\gamma$--ray emission, as indicated by the
slope measured by {\it Fermi}, is related to the slope of the
electron energy distribution in its high energy part.
It is then associated to the slope of the high energy synchrotron emission.

\item
The bolometric power of our sources is always dominated by the $\gamma$--ray flux.
The ratio between the high to low energy emission humps ($L_{\rm C}/L_{\rm S}$)
is directly related to the ratio between the radiation to magnetic energy
density  $U^\prime_{\rm r}/U^\prime_{\rm B}$.
If the dissipation region is within the BLR, 
our assumption of $R_{\rm BLR}=10^{17} L_{\rm d,45}^{1/2}$ cm gives
\begin{equation}
{U^\prime_{\rm r} \over U^\prime_{\rm B}} \, =\, {L_{\rm C}\over L_{\rm S}}\, \to
U^\prime_{\rm B} \, =\, {L_{\rm S}\over L_{\rm C}} \, {\Gamma^2 \over 12\pi}\,
\to B=\Gamma\left( {2L_{\rm s}\over 3 L_{\rm C} } \right)^{1/2}
\label{b}
\end{equation}
where we have assumed that $U^\prime_{\rm r}\approx U^\prime_{\rm BLR}$.

\item
The peak of the high energy emission ($\nu_{\rm C}$)
is believed to be produced 
by the scattering of the line photons (mainly hydrogen Lyman--$\alpha$) with 
electrons at the break of the particle distribution ($\gamma_{\rm peak}$).
Its observed frequency, $\nu_{\rm C}$, is
\begin{equation}
\nu_{\rm C} \, \sim \, {2 \over 1+z} \nu_{Ly\alpha} \Gamma\delta \gamma_{\rm peak}^2 
\label{nuc}
\end{equation}
A steep ($\alpha>1$) spectrum indicates a peak at energies below 100 MeV,
and this constrains $\Gamma\delta \gamma_{\rm peak}^2$.
For our sources, the scattering process leading to the formation of
the high energy peak occurs always in the Thomson regime.

\item
The strength of the SSC relative to the EC emission depends
on the ratio between the synchrotron over the external radiation energy densities,
as measured in the comoving frame, $U^\prime_{\rm s}/U^\prime_{\rm ext}$.
Within the BLR, $U^\prime_{\rm ext}$ depends only on $\Gamma^2$,
while $U^\prime_{\rm s}$ depends on the injected power, the 
size of the emission, and the magnetic field.
The larger the magnetic field, the larger the SSC component.
The shape of the EC and SSC emission is different:
besides the fact that the seed photon distributions are different,
we have that the flux at a given X--ray frequency
is made by electron of very different energies,
thus belonging to a different part of the electron distribution.
In this respect, the low frequency X--ray data of very hard X--ray spectra
are the most constraining,
since in these cases the (softer) SSC component must not exceed what observed.
This limits the magnetic field, the injected power (as measured in the comoving
frame) and the size.
Conversely, a relatively soft spectrum (but still rising, in $\nu F_\nu$) 
indicates a SSC origin, and this constrains the combination of $B$, $r_{\rm diss}$ 
and $P^\prime_{\rm i}$
even more.

\item
In our powerful blazars the radiative cooling rate is almost
complete, namely even low energy electrons cool in a 
dynamical timescale $r_{\rm diss}/c$.
We call $\gamma_{\rm cool}$  
the random Lorentz factor of those electrons halving their energies
in a timescale $r_{\rm diss}/c$.
Note that $\gamma_{\rm cool}$ 
is rarely very close to unity, but more often a few).
Therefore the corresponding emitting particle distribution is weakly
dependent of the low energy spectral slope, $s_1$, of the injected
electron distribution.

\item 
In several sources we see clear signs of thermal emission in the
optical--UV part of the spectrum.
If associated to the flux produced by a standard 
multi--colour accretion disk, we can estimate both the mass
of the black hole and the total disk luminosity.
The temperature profile of a standard disk emitting locally
as a black--body is (e.g. Frank, King \& Raine 2002)
\begin{equation}
T^4 \, =\, {  3 R_{\rm S}  L_{\rm d }  \over 16 \pi\eta\sigma_{\rm MB} R^3 }  
\left[ 1- \left( {3 R_{\rm S} \over  R}\right)^{1/2} \right]   
\end{equation}
where $L_{\rm d}=\eta \dot M c^2$ is the bolometric disk luminosity and
$R_{\rm S}$ is the \sc\ radius.
The maximum temperature (and hence the peak of the $\nu F_\nu$ disk spectrum)
occurs at $R\sim 5 R_{\rm S}$ and scales as
$T_{\rm max}\propto (L_{\rm d}/L_{\rm Edd})^{1/4}M^{-1/4}$.
The level of the emission gives $L_{\rm d}$ 
[that of course scales as $(L_{\rm d}/L_{\rm Edd})\, M$].
Therefore, for a good optical--UV coverage, and when the synchrotron
emission is weaker than the thermal emission, we can derive both
the black hole mass and the accretion luminosity.
Given a value of the efficiency $\eta$, this translates in 
the mass accretion rate.

\item
The one--zone homogeneous model here adopted is aimed to explain the 
bulk of the emission, and necessarily requires a compact source,
self--absorbed (for synchrotron) at $\sim 10^{12}$ Hz.
The flux at radio frequencies must be produced further out in the jet.
Radio data, therefore, are not directly constraining the model.
Indirectly though, they can suggest a sort of continuity between the 
level of the radio emission and what the model predicts at higher
frequencies.

\end{itemize}

\begin{table*} 
\centering
\begin{tabular}{llllllllllllll}
\hline
\hline
Name   &$z$ &$R_{\rm diss}$ &$M$ &$R_{\rm BLR}$ &$P^\prime_{\rm i}$ &$L_{\rm d}$ &$B$ &$\Gamma$ &$\theta_{\rm v}$
    &$\gamma_{\rm b}$ &$\gamma_{\rm max}$ &$s_1$  &$s_2$  \\
~[1]      &[2] &[3] &[4] &[5] &[6] &[7] &[8] &[9] &[10] &[11] &[12] &[13] &[14]  \\
\hline   
0048--071       &1.975  &210 (700)   &1e9   &474 &0.025 &22.5 (0.15) &2.4   &15.3 &3  &400   &7e3   &1   &2.7  \\ 
0202--17        &1.74   &300 (1000)  &1e9   &671 &0.03  &45 (0.3)    &2.4   &15   &3  &300   &5e3   &1   &3.1  \\ 
0215+015        &1.715  &900 (1500)  &2e9   &548 &0.04  &30 (0.1)    &1.1   &13   &3  &2.5e3 &6e3   &--1 &3.5  \\
0227--369       &2.115  &420 (700)   &2e9   &547 &0.08  &30 (0.1)    &1.5   &14   &3  &200   &5e3   &0   &3.1  \\
0235+164        &0.94   &132 (440)   &1e9   &212 &0.042 &4.5 (0.03)  &2.9   &12.1 &3  &400   &2.7e3 &--1 &2.1   \\
0347--211       &2.944  &750 (500)   &5e9   &866 &0.12  &75 (0.1)    &1.5   &12.9 &3  &500   &3e3   &--1 &3.0   \\
0426--380       &1.112  &156 (1300)  &4e8   &600 &0.018 &36 (0.6)    &1.7   &13   &3  &300   &6e3   &--1 &2.4   \\
0454--234       &1.003  &338 (450)   &2.5e9 &433 &0.027 &18.8 (0.05) &3     &12.2 &3  &330   &4e3   &--1 &2.4   \\
0528+134$_{93}$ &2.04   &540 (1800)  &1e9   &866 &1.1   &75 (0.5)    &1.7   &15   &3  &150   &4e3   &--1 &2.3   \\
0528+134$_{95}$ &2.04   &315 (1050)  &1e9   &866 &0.45  &75 (0.5)    &3.5   &13   &3  &160   &5e3   &--1 &2.3   \\
0528+134$_{97}$ &2.04   &300 (1000)  &1e9   &866 &0.1   &75 (0.5)    &3.6   &13   &3  &120   &3e3   &--1 &3.0   \\
0528+134$_{08}$ &2.04   &420 (1400)  &1e9   &866 &0.13  &75 (0.5)    &2.6   &13   &3  &150   &3e3   &--1 &2.8   \\
0820+560        &1.417  &261 (580)   &1.5e9 &581 &0.023 &34 (0.15)   &3.1   &13.9 &3  &220   &3e3   &0   &3.4   \\
0917--449$_F$   &2.1899 &900 (500)   &6e9   &1341 &0.1  &180 (0.2)   &1.95  &12.9 &3  &50    &4e3   &--1 &2.6   \\
0917--449$_E$   &2.1899 &900 (500)   &6e9   &1341 &0.1  &180 (0.2)   &1.95  &12.9 &3  &30    &4e3   &--1 &2    \\
1013+054        &1.713  &252 (420)   &2e9   &300 &0.036 &9 (0.03)    &1.7   &11.8 &3  &500   &3e3   &1   &2.4   \\
1329--049       &2.15   &450 (1000)  &1.5e9 &822 &0.07  &67.5 (0.3)  &1.4   &15   &3  &300   &5e3   &1   &3.3  \\
1454--354$_{F_1}$ &1.424 &150 (250)  &2e9   &671 &0.05  &45 (0.15)   &4.9   &15   &3  &450   &3e3   &--1 &2.4  \\
1454--354$_{F_2}$ &1.424 &240 (400)  &2e9   &671 &0.025 &45 (0.15)   &3.9   &11.5 &3  &100   &3e3   &--1 &2.0  \\
1454--354$_{F_3}$ &1.424 &150 (250)  &2e9   &671 &0.25  &45 (0.15)   &2.0   &20.  &3  &1e3   &4e3   &--1 &2.0   \\
1502+106        &1.839  &450 (500)   &3e9   &764 &0.16  &58.5 (0.13) &2.8   &12.9 &3  &600   &4e3   &--1 &2.1   \\
1520+319        &1.487  &1500 (2000) &2.5e9 &237 &0.04  &5.6 (0.015) &0.06  &15   &3  &2e3   &3e4   &0.8 &2.6   \\
1551+130        &1.308  &330 (1100)  &1e9   &755 &0.02  &57 (0.38)   &2.0   &13   &3  &200   &6e3   &--1 &2.4   \\
1633+382        &1.814  &750 (500)   &5e9   &866 &0.07  &75 (0.1)    &1.5   &12.9 &3  &230   &6e3   &0   &2.9   \\
2023--077       &1.388  &378 (420)   &3e9   &474 &0.07  &22.5 (0.05) &1.8   &11.8 &3  &350   &4e3   &0   &2.6   \\ 
2052--47        &1.4910 &210 (700)   &1e9   &612 &0.045 &37.5 (0.25) &2.6   &13   &3  &100   &7e3   &--1 &3.0   \\
2227--088       &1.5595 &211 (470)   &1.5e9 &497 &0.06  &24.8 (0.11) &3.3   &12   &3  &200   &5e3   &0.5 &3.2   \\
2251+158$_{00}$ &0.859  &300 (1000)  &1e9   &548 &0.05  &30 (0.2)    &3.3   &13   &3  &40    &7e3   &0.5 &2.4   \\
2251+158$_{05}$ &0.859  &120 (400)   &1e9   &548 &0.15  &30 (0.2)    &15.5  &11.5 &3  &600   &1.5e3 &0   &3.5   \\
2251+158$_{07}$ &0.859  &270 (900)   &1e9   &548 &0.22  &30 (0.2)    &3.6   &13   &3  &20    &3.5e3 &1.6 &1.6    \\
2251+158$_{08}$ &0.859  &240 (800)   &1e9   &548 &0.14  &30 (0.2)    &4.1   &13   &3  &250   &4e3   &1   &2.7   \\
2325+093        &1.843  &420 (1400)  &1e9   &671 &0.08  &45 (0.3)    &1.6   &16   &3  &190   &5e3   &0   &3.5   \\
\hline
\hline 
\end{tabular}
\vskip 0.4 true cm
\caption{List of parameters used to construct the theoretical SED.
Subscripts refer to the year of the corresponding SED
or to an EGRET ($E$) state as opposed to a {\it ``Fermi"} ($F$) state.
Col. [1]: name;
Col. [2]: redshift;
Col. [3]: dissipation radius in units of $10^{15}$ cm and (in parenthesis) in units of $R_{\rm S}$;
Col. [4]: black hole mass in solar masses;
Col. [5]: size of the BLR in units of $10^{15}$ cm;
Col. [6]: power injected in the blob calculated in the comoving frame, in units of $10^{45}$ erg s$^{-1}$; 
Col. [7]: accretion disk luminosity in units of $10^{45}$ erg s$^{-1}$ and
        (in parenthesis) in units of $L_{\rm Edd}$;
Col. [8]: magnetic field in Gauss;
Col. [9]: bulk Lorentz factor at $R_{\rm diss}$;
Col. [10]: viewing angle in degrees;
Col. [11] and [12]: break and maximum random Lorentz factors of the injected electrons;
Col. [13] and [14]: slopes of the injected electron distribution [$Q(\gamma)$] below and above $\gamma_{\rm b}$;
For all cases the X--ray corona luminosity $L_X=0.3 L_{\rm d}$.
Its spectral shape is assumed to be $\propto \nu^{-1} \exp(-h\nu/150~{\rm keV})$.
}
\label{para}
\end{table*}

\begin{table} 
\centering
\begin{tabular}{lllll}
\hline
\hline
Name   &$\log P_{\rm r}$ &$\log P_{\rm B}$ &$\log P_{\rm e}$ &$\log P_{\rm p}$ \\
\hline   
PKS 0048--071   &45.75 &45.35 &44.69 &47.10   \\
PKS 0202--17    &45.80 &45.65 &44.80 &47.31   \\
PKS 0215+015    &45.81 &45.78 &44.86 &46.09    \\
PKS 0227--369   &46.18 &45.49 &44.97 &47.34   \\
AO  0235+164    &45.78 &44.97 &44.61 &46.60   \\
PKS 0347--211   &46.30 &45.91 &44.55 &46.92    \\
PKS 0426--380   &45.48 &44.67 &44.30 &46.36   \\
PKS 0454--234   &45.60 &45.80 &44.16 &46.40   \\
PKS 0528+134$_{93}$  &46.32 &45.88 &45.12 &47.52   \\
PKS 0528+134$_{95}$  &46.19 &45.88 &45.22 &47.61   \\
PKS 0528+134$_{97}$  &46.87 &45.88 &45.59 &47.94   \\
PKS 0528+134$_{08}$  &47.39 &45.86 &45.87 &48.31   \\
TXS 0820+560    &45.60 &45.75 &44.57 &46.85  \\
TXS 0917+449$_F$  &46.20 &46.29 &45.00 &47.57   \\
TXS 0917+449$_E$  &46.20 &46.29 &44.87 &47.43   \\
TXS 1013+054    &45.68 &45.02 &44.66 &46.99   \\
PKS 1329--049   &46.18 &45.53 &45.07 &47.65   \\
PKS 1454--354$_{F_1}$ &46.04 &45.65 &44.59 &46.80    \\
PKS 1454--354$_{F_2}$ &45.50 &45.65 &44.38 &46.53    \\
PKS 1454--354$_{F_3}$ &47.01 &45.13 &45.20 &47.47   \\
PKS 1502+106    &46.43 &46.12 &44.63 &46.91   \\
TXS 1520+319   &45.91 &43.84 &45.22 &46.50   \\
PKS 1551+130    &45.53 &45.45 &44.21 &46.48   \\
TXS 1633+382    &46.06 &45.91 &44.60 &47.05    \\
PKS 2023--077   &45.98 &45.41 &44.65 &46.87    \\
PKS 2052-47     &45.84 &45.27 &45.01 &47.24   \\
PKS 2227--088   &45.89 &45.40 &45.07 &47.29    \\
PKS 2251+158$_{00}$  &45.87 &45.78 &45.11 &47.56    \\
PKS 2251+158$_{05}$  &46.24 &46.34 &44.84 &47.12    \\
PKS 2251+158$_{07}$  &46.53 &45.78 &45.61 &48.24    \\
PKS 2251+158$_{08}$  &46.33 &45.78 &45.43 &47.87    \\
PKS 2325+093     &46.30 &45.65 &45.06 &47.51  \\
\hline
\hline 
\end{tabular}
\vskip 0.4 true cm
\caption{
Jet power in the form of radiation, Poynting flux,
bulk motion of electrons and protons (assuming one proton
per emitting electron).
Subscripts refer to the year of the corresponding SED
or to an EGRET ($E$) state as opposed to a {\it ``Fermi"} ($F$)
state.}
\label{powers}
\end{table}
%

\begin{figure}
\vskip -0.7cm
\hskip -0.9cm
\psfig{figure=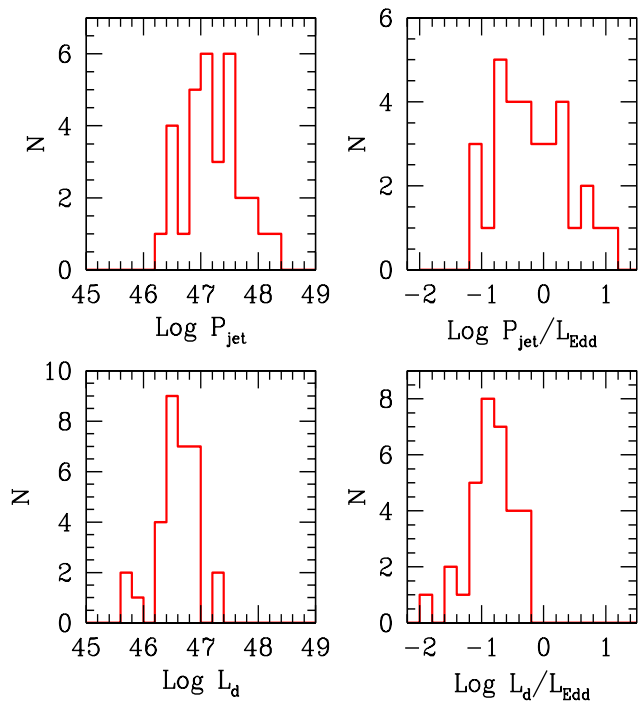,width=10cm,height=11cm}
\vskip -0.6 cm
\caption{Histograms of the total jet power and the
accretion disk luminosity, both in cgs units (left)
and in Eddington units (right).
}
\label{isto1}  
\end{figure}

\begin{figure}
\vskip -0.7cm
\hskip -0.9cm
\psfig{figure=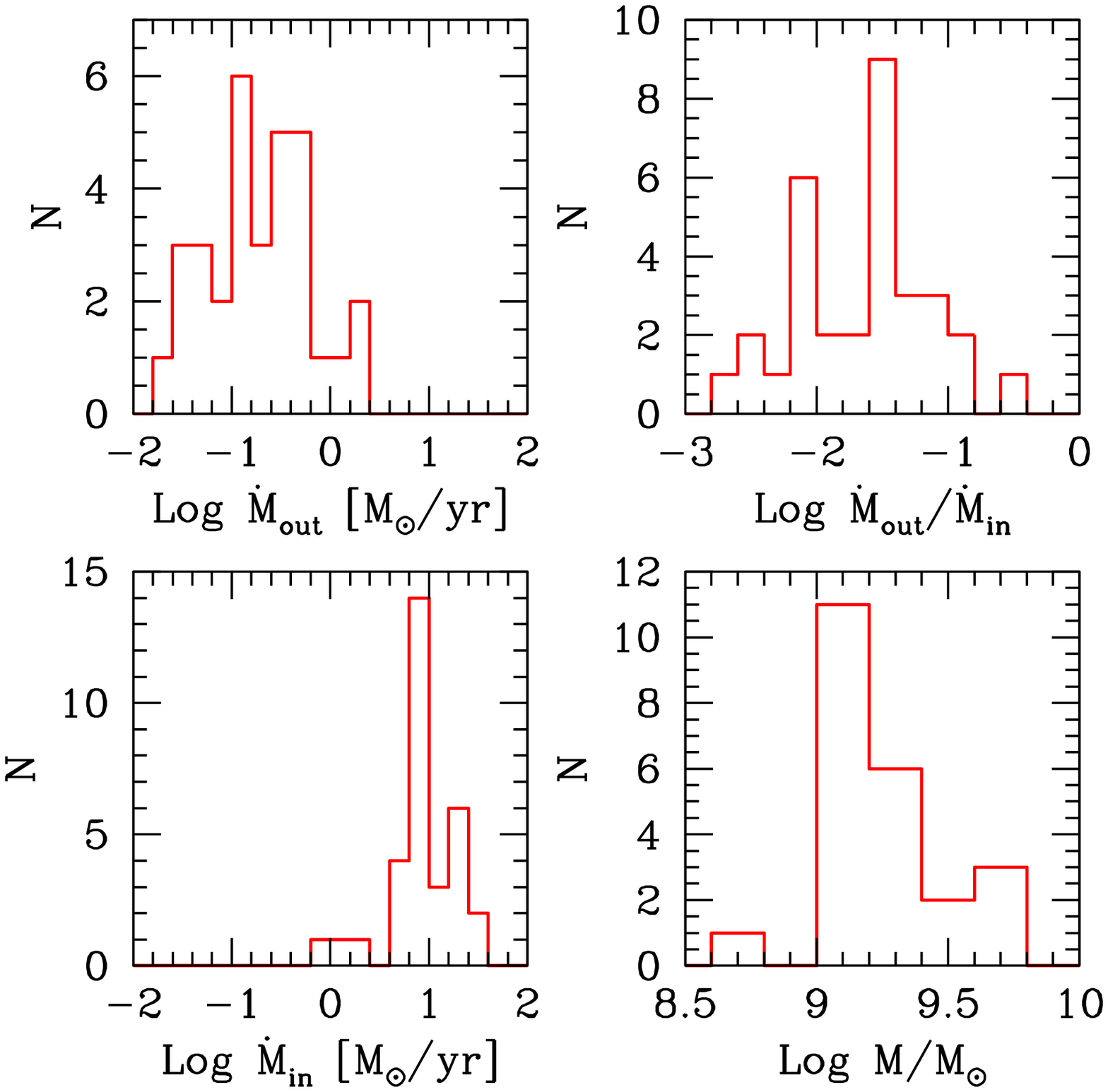,width=10cm,height=11cm}
\vskip -0.6 cm
\caption{Histograms of the outflowing (top left) mass
rate; the accretion mass rate (bottom left) and
their ratio (top right).
On the bottom right we show the distribution
of black hole masses found in this paper.
}
\label{isto2}  
\end{figure}


\section{Results}

The SED of our sources, together with the best fitting model, 
are shown in Figs. \ref{f1} -- \ref{f7}.
In Tab. \ref{xrt} and Tab. \ref{uvot} we list the result of our
XRT and UVOT analysis for the 17 blazars with {\it Swift} data.
In Tab. \ref{para} we list the parameters used to compute 
the theoretical SEDs and in Tab. \ref{powers} we list the
the power carried by the jet in the form of radiation, electrons, magnetic field
and protons (assuming one proton per emitting electron).
All optical--UV fluxes shown in Fig. \ref{f1} -- \ref{f7} have been
de--reddened according to the Galactic $A_{\rm V}$ given in the
NED database and reported in Tab. \ref{uvot}, but not for the (possible) 
absorption in the host galaxy
nor for Ly$\alpha$ absorption (line and edge) due to intervening matter
along the line of sight. 
According both to theoretical consideration (see Madau, Haardt \& Rees 1999)
and optical--UV spectra of high redshift quasars (e.g. Hook et al. 2003) this kind
of absorption, in quasars, should be at most marginal.
This is an important point, since it allows us to consider the bluer 
photometric points of UVOT as a relatively good estimate of the 
intrinsic flux of the source.
Inspection of the optical--UV SED shown in Figs. \ref{f1} -- \ref{f7} strongly 
suggests that in some cases this emission is due to the accretion disk.
This is when there is a peak in the $\nu F_\nu$ optical--UV
spectrum with an exponential decline. 
This cannot be the synchrotron peak, since
the steep spectral index implied by the steep $\gamma$--ray energy emission 
(bow--ties) is incompatible with this interpretation.

This fact, together with the theoretical considerations
mentioned in the previous section, allow us to consider 
the optical--UV emission of nearly half of our blazars
as due to the accretion disk emission.
Among the sources with UVOT data exceptions are 
0215+015, 0235+164, 0454--234, 0528+134,
1454--354, 2251--158 and 2325+093. 
For these sources either the non--thermal continuum
dominates the flux, or the quality of the data is not sufficient
to discriminate between the thermal and the non--thermal processes.
For them, the black hole mass and disk luminosity has been
chosen (when summed to the non--thermal emission)
not to overproduce the observed flux.

\subsection{Black hole masses, accretion and outflow rates}

The derived black hole masses are in the range (1--6)$\times 10^9M_\odot$
(Fig. \ref{isto2}),
and the disk luminosities, in Eddington units, cluster 
around $L_{\rm d}/L_{\rm Edd}\sim 0.1$--0.3 (see Fig. \ref{isto1}).
Fig. \ref{isto2} shows the accretion and outflow mass rates (in units
of $M_\odot$ yr$^{-1}$ and their ratio. 
All our sources requires more than 1 $M_\odot$ yr$^{-1}$ 
of accreting mass, while the outflow rate is a factor
0.01--0.1 smaller.
This can be understood in simple terms, considering that 
in matter dominated jets $P_{\rm j} =\Gamma \dot M_{\rm out} c^2$ and then
\begin{equation}
{\dot M_{\rm out} \over \dot M_{\rm accr}} \, =\, 
{\eta \over \Gamma}\, {P_{\rm j}\over L_{\rm d}} \, \approx  \,
{\eta_{-1} \over \Gamma_1}\,
10^{-2} {P_{\rm j}\over L_{\rm d}}
\end{equation}

\subsection{Jet power}

Table \ref{powers} lists the power carried by the jet in the form of
radiation ($P_{\rm r}$), magnetic field ($P_{\rm B}$), electrons
($P_{\rm e}$) and protons ($P_{\rm p}$, assuming one proton
per emitting electron). 
All the powers are calculated as
\begin{equation}
P_i  \, =\, \pi r^2 \Gamma^2\beta c \, U_i
\end{equation}
where $U_i$ is the energy density of the $i$ component.

The total power carried by the jet is often larger than the disk 
luminosity, and sometimes exceeds the Eddington luminosity
(see Fig. \ref{isto1} and Fig. \ref{all}).
In Fig. \ref{all} we compare $P_{\rm jet}$ as a function of $L_{\rm d}$
for our sources with the FSRQs and BL Lacs analysed in  CG08
and with the FSRQs analysed
in Maraschi et al. (2008, hereafter M08).
As can be seen, the blazars analysed in CG08 and M08 tend to have 
larger jet powers with respect to 
disk luminosities than the blazars presented here.
This can be partly understood considering that a large fraction
of the blazars in CG08 were EGRET sources, and the $\gamma$--ray data
correspond to high state of the sources.
In our sample, instead, the $\gamma$--ray luminosity, derived by
the {\it average} $\gamma$--ray flux seen by {\it Fermi}, is
considerably smaller than the maximum values seen by EGRET
(see Fig. \ref{f0528} as illustrative example).
The smaller electromagnetic power output implies that also the 
kinetic power carried by the jet is smaller.
The blazars considered in M08 (see also Sambruna et al. 2007
and Ghisellini 2009), instead, lack $\gamma$--ray data,
and the estimates were based on fitting the SED up to the {\it Swift}--XRT
energy range. 
Several of them showed a very hard X--ray spectrum, suggesting a peak
in the 1--10 MeV region of the spectrum.
MeV blazars of this kind, peaking at energies significantly smaller
than 100 MeV, may not be easily observable by {\it Fermi},
and could be among the most luminous blazars\footnote{The recently published list
of blazars detected in the 15--55 keV band in the 3--years survey of 
the Burst Alert Telescope (BAT) onboard {\it Swift} indeed includes very 
powerful blazars (Ajello et al. 2009), up to redshifts $z=4.1$.
}  
(according to the 
blazar sequence, Fossati et al. 1998).
The jet powers calculated here may not
correspond to the ``very high states" of blazars as seen by EGRET.
They are therefore more representative of a typical blazar state,
but we have to remember that when constructing a list of variable sources
(even with a more sensitive instrument) high states are always 
over--represented.
What is confirmed with these new data is that, at least occasionally, 
the jet is more powerful than $L_{\rm d}$.
 
We can also compare the fraction of the total jet power carried
in radiation ($\epsilon_{\rm r}$), magnetic field ($\epsilon_{\rm B}$), 
and electrons ($\epsilon_{\rm e}$) with the sample of blazars 
studied in CG08.
This is done in Fig. \ref{epsilon}, showing a rough consistency
with the blazars in CG08.

\subsection{Emission mechanisms}

The location of the dissipation region is almost always
within the BLR (exceptions are 0215+015 and 1520+319), that
provides most of the seed photons scattered at high frequencies.
The main emission process are then synchrotron and thermal emission
from the accretion disk for the low
frequency parts, and External Compton for the hard X--rays and the $\gamma$--ray
part of the spectrum.
The X--ray corona and the SSC flux marginally contribute to the soft X--ray part of
the SED.
The IR radiation reprocessed by the torus can in some case dominate
the far IR flux, but the scarcity of data in this frequency window
does not allow to fully assess its importance.
Note that when $R_{\rm diss}<R_{\rm BLR}$, the overall non--thermal
emission is rather insensitive to the presence/absence of the IR torus,
since the bulk of the seed photons are provided by the broad lines.
However, for 0215+015 and 1520+319, the location of the dissipation
region is beyond the BLR, and for these two sources the external
Compton process with the IR radiation of the torus is crucial
to explain their SED.
Since their dissipation region is beyond the BLR, the
corresponding energy densities in radiation and fields are smaller than 
in the other sources, as can be seen in the top panel of Fig. \ref{gpeak}.
Furthermore, the smaller $U_{\rm r}+U_{\rm B}$ implies a slower cooling,
and thus a larger $\gamma_{\rm cool}$ (bottom panel of Fig. \ref{gpeak}).

\subsection{Particle distribution}

The lower end of the emitting particle distribution
is usually the bottleneck to reliably calculate
how many particle the jet carries, and hence its power.
This is because low energy electrons emit, by synchrotron,
in the self--absorbed, not visible, regime, and, by SSC,
they contribute in a region of the SED usually dominated
by synchrotron or by the accretion disk.
There are however several cases where the X--ray data
can be convincingly fitted by the EC mechanism, and
in these cases the IC radiation produced  by these low energy
electrons is dominating the observed flux.
These are the cases where the X--ray spectral shape
is very hard, especially at lower frequencies, excluding SSC as 
the main contributor to the observed flux.
As examples, see the {\it Beppo}SAX spectrum of 2251--158 in the
year 2000, or the XRT spectrum of 2325+093.
These are the cases where the $N(\gamma)$ distribution
can be reliably determined even in its low energy part.
Theoretically, as an outcome of the model assumption,
we can also calculate the random Lorentz factor of
the electrons cooling in one light crossing time.
This depends on the amount of radiative cooling,
in turn depending on the bulk Lorentz factor,
the magnetic field, and the amount of external radiation.
Fig. \ref{gpeak} shows the values of $\gamma_{\rm cool}$
(as a function of the peak energy $\gamma_{\rm peak}$) 
for our sources.
Since we are dealing with very powerful blazars, where
the cooling is strong, the derived values are particularly
small, of the order of 1--10.
This ensure that $N(\gamma)\propto \gamma^{-2}$ between 
$\gamma_{\rm cool}$ and $\gamma_{\rm b}$.

For all our sources the importance of the $\gamma$--$\gamma\to e^\pm$ process
(which is included in our model) is very modest, and does not 
influence the observed spectrum.

\begin{figure}
\vskip -0.5cm
\psfig{figure=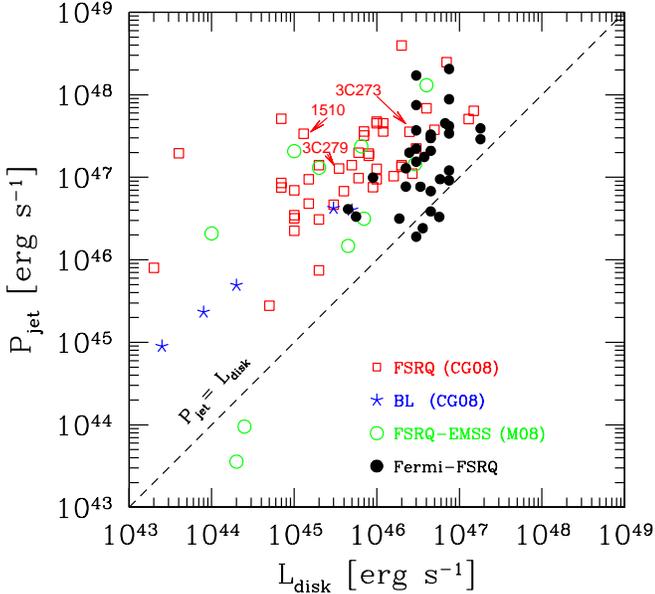,width=9cm,height=9cm}
\vskip -0.5 cm
\caption{The jet power as a function
of the accretion disk luminosity of different
blazar samples.
Black filled circles: our sources;
empty squares: FSRQs analysed in Celotti \& Ghisellini (2008, CG08));
stars: BL Lac objects analysed in CG08;
empty circles: FSRQs analysed in Maraschi et al. (2008, M08).
3C 273 and PKS 1510--089 are labelled, to show their position 
in this plane (see text).
}
\label{all}  
\end{figure}

\begin{figure}
\vskip -0.5cm
\hskip -0.5 cm
\psfig{figure=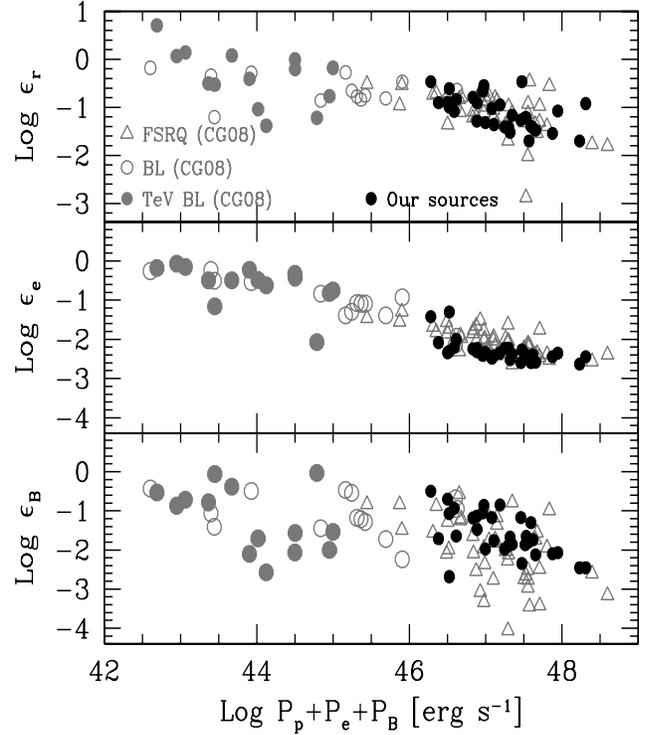,width=9.7cm,height=11.5cm}
\vskip -0.5 cm
\caption{
The fraction of $L_{\rm jet}$ radiated 
($\epsilon_{\rm r}$, top panel), in relativistic leptons 
($\epsilon_{\rm e}$, mid panel) and in magnetic fields 
($\epsilon_B$, bottom panel) as functions of 
$P_{\rm jet}=P{\rm p}+P_{\rm e}+P_{\rm B}$. 
Our sources (black circles) are compared to the FSRQs (grey triangles),
BL Lacs (grey empty circles) and TeV BL Lacs (grey filled circles)
of the blazar sample studied in CG8.
}
\label{epsilon}  
\end{figure}

\begin{figure}
\vskip -0.5cm
\hskip -1.7 cm
\psfig{figure=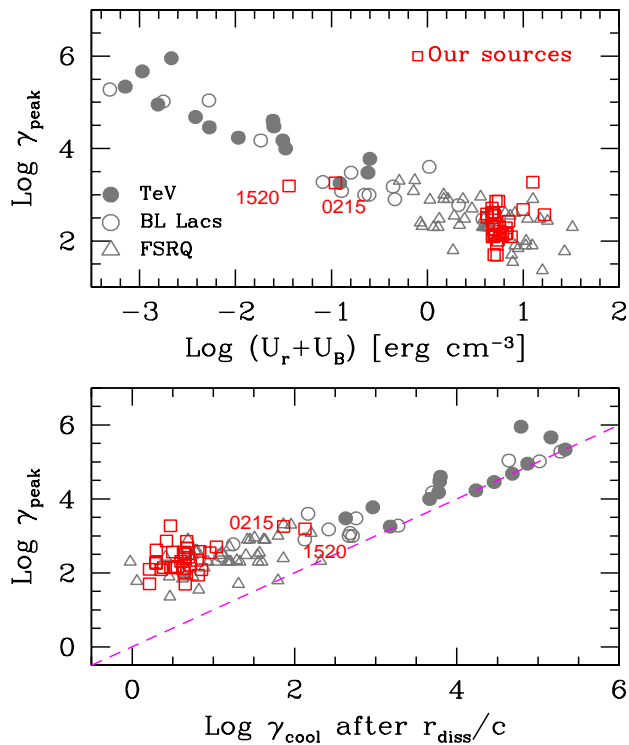,width=11.5cm,height=11cm}
\vskip -0.5 cm
\caption{Top: $\gamma_{\rm peak}$ vs $U_B+U^\prime_{\rm r}$.
Bottom:   $\gamma_{\rm peak}$ vs $\gamma_{\rm cool}$
calculated after one light crossing time.
The dashed line indicate equality.
See text for details.
In both panels the  grey symbols are the blazars studied
in Celotti \& Ghisellini (2008).
The two sources outside the cluster of the other ones are 
PKS 0215+015 and TXS 1520+319, as labelled. 
These are the only two sources with $R_{\rm diss}>R_{\rm BLR}$ and
this implies a reduced energy density (in radiation
and magnetic field) and therefore a larger $\gamma_{\rm cool}$.
}
\label{gpeak}  
\end{figure}

\subsection{Comparison with low power jets}

Fig. \ref{gpeak} shows the value of the peak energy $\gamma_{\rm peak}$
(namely the random Lorentz factor of the electrons producing 
most of the emission at the two peaks of the SED) as a function
of the total energy density (as seen in the comoving frame; upper panel)
and as a function of $\gamma_{\rm cool}$ (calculated after one light crossing time 
$r_{\rm diss}/c$; bottom panel).
We have reported, for comparisons, the values obtained by CG08
for a sample of blazars including both powerful FSRQs and low
power BL Lacs.
As expected, the values for our blazar are consistent with those
previously found by CG08 for powerful FSRQs.
The bottom panel of Fig. \ref{gpeak} is a different version of the 
blazar sequence as originally proposed in Ghisellini et al. (1998) and
shown in the upper panel.
It suggests that $\gamma_{\rm peak}$ becomes equal to $\gamma_{\rm cool}$ 
when the cooling is weak, but it must be larger than $\gamma_{\rm cool}$ 
when the cooling is fast.
Therefore the position of the peaks of the SED of strong FSRQs 
is controlled by the injection
function $Q(\gamma)$ and its break at $\gamma_{\rm b}$, that
is proportional to $\gamma_{\rm peak}$ (the two energies are not
exactly equal because of the assumed smoothly curved injection
function, see GT09).
In the slow cooling regime, instead (low power BL Lacs),
$\gamma_{\rm b}$ becomes unimportant, and the particle distribution
self--adjust itself.
The new feature added here is the fact that in this case (i.e. slow cooling) 
the value of $\gamma_{\rm peak}=\gamma_{\rm cool}$ depends also on the
size of the source, and not only on the total energy density.

\subsection{Multi--states of specific sources}

\noindent
{\bf PKS 0528+134 ---}
Fig. \ref{f3} shows 4 states of this source, one of the best
studied by EGRET, that was able to detect it at very different levels
of $\gamma$--ray activity 
(see also Mukherjee et al. 1999 for a detailed analysis
of the different states during the EGRET era).
Keeping the black hole mass and the accretion rate fixed (so that also $L_{\rm d} $ 
and $R_{\rm BLR}$ are fixed), we tried to interpret the 4 different states changing 
a minimum number of parameters.
Also the bulk Lorentz factor need not to change.
Indeed, the main changing parameters are i) $R_{\rm diss}$, implying a change of the 
magnetic field in the dissipation region (larger field at closer distances, according
to a constant $P_{\rm B}$, as listed in Tab. \ref{powers}), 
and ii) the power $P^\prime_{\rm i}$ injected into the source in the form 
of relativistic electrons.
The change in $P^\prime_{\rm i}$ is the main responsible for the changes in 
the jet bulk kinetic power (factor $\sim$5), that are sufficient to explain 
the rather dramatic changes in the observed SED.
We had to change also the high energy slope ($s_2$) of the injected
electron distribution, and (but by a modest factor) $\gamma_1$ and $\gamma_{\rm max}$.
The changes of the magnetic field (factor $\sim$2) are responsible for 
a very different contribution of the SSC flux in the X--ray band,
dominated by this component in the SEDs of 1993 and 1995.

\vskip 0.3 cm
\noindent
{\bf TXS 0917+449 = RGB 0920+446 ---}
In Fig. \ref{f3bis} we show 2 models for the SED of this source, corresponding to 
the combined {\it Swift} and {\it Fermi} data, and also to the
old EGRET ones (for which there is a paucity of data).
Therefore the two models do not describe two different observed SEDs, 
but can illustrate how the model parameters have to change if only the 
$\gamma$--ray emission changes.
In this case we have fixed, besides the black hole mass and accretion rate, also the 
location of the dissipation region, the magnetic field and the injected power.
We only changed the high energy slope of the injected electron distribution ($s_2$, from
2.6 to 2), and $\gamma_1$ (from 50 to 30).
This changes are responsible for the (small) changes of the
total bulk kinetic power of the jet.

\vskip 0.3 cm
\noindent
{\bf PKS 1454--354 ---}
Fig. \ref{f4} shows 3 SEDs for this source, corresponding to three states
already observed by {\it Fermi} and discussed in Abdo et al. (2009c).
Fig. \ref{f4} reports the $\gamma$--ray data  of the flare, of the ``low" state
and of the intermediate $\gamma$--ray state that corresponds to the average flux
reported in Abdo et al. (2009a).
{\it Swift} data corresponds to Jan and Sept 2008.
As for the other sources, we have tried to model the 3 states of the source
assuming the same black hole mass and disk luminosity (hence the same $R_{\rm BLR}$).
Values of $R_{\rm diss}$ are constrained by the observed short variability timescale
of the $\gamma$--ray flux (Abdo et al. 2009c).
We can reproduce the different source states by changing $R_{\rm diss}$ (factor 2),
$P^\prime_{\rm i}$ (factor 5), $\Gamma$ (almost a factor 2) and $B$ (factor $\sim$2.5).
In addition, we had to adjust $\gamma_1$ and $s_2$.
These changes of the physical parameters of the jet imply different jet powers 
(factor $\sim$2 for $P_{\rm B}$ and $\sim$10 for the bulk kinetic jet power).

\vskip 0.3 cm
\noindent
{\bf 2251--158 = 3C 454.3 ---}
Fig. \ref{f7} shows 4 states of this source, one of the brightest {\it AGILE}
blazars, and also detected, besides by {\it Fermi}, also by EGRET.
The first 3 SEDs have been discussed by Ghisellini et al. (2007), with the
difference that at that time only an integrated {\it AGILE} flux was known,
while now (Vercellone et al. 2009) the spectral analysis of the {\it AGILE} data
yielded a spectrum {\it rising} in $\nu F_\nu$.
The fourth corresponds to the {\it Fermi} observation (with a flux 
slightly smaller than {\it AGILE} but with a steep slope).
This blazars shows a rather dramatic variability at all frequencies,
most notably in the optical band in 1995
(Fuhrmann et al. 2006; 
Villata et al. 2006; 
Giommi et al. 2006; 
Pian et al. 2006). 
Comparing the SED of the 2005 flare with the SED of 2000, the IR--optical flux
increased by two orders of magnitude, while the hard X--rays flux increased by a 
a factor $\sim$10.
Unfortunately, for the SED in 2005 there are no information
about the $\gamma$--ray flux.
Katarzynski \& Ghisellini (2007) interpreted this behaviour (namely, more 
variability amplitude in the optical than at X--rays) suggesting that
the main change was due to the location of the dissipation region in the 
jet, more internal in 2005. 
They showed that even with a constant jet power, and a constant
bolometric luminosity, dramatic variations in specific bands were 
possible.
The reason is that a more internal dissipation region may
correspond to a larger magnetic field and a smaller bulk Lorentz
factor (i.e. the jet may still be accelerating): the combination
of these two key factors yields a larger magnetic energy density
and a smaller external radiation field (since in the comoving frame 
it is less amplified by the smaller $\Gamma$).
This in turn implies more synchrotron and less EC radiation.
These ideas were later confirmed (Ghisellini et al. 2007)
by the first  {\it AGILE} observations, that caught the source 
in a bright $\gamma$--ray state and in a relatively faint optical state 
(compared to 1995). 
We here follow the same ideas,
adopting a smaller $R_{\rm diss}$ and $\Gamma$ for the 2005 state
(see Tab. \ref{para}).
The SED corresponding to the {\it Fermi/Swift} observations 
in 2008 confirms this scenario: the source is still much brighter
in $\gamma$--rays than revealed by EGRET, but it is not exceptionally 
strong in optical.
What is notably different, for the 2007 and 2008 states, is the slope of the
$\gamma$--ray spectrum, which is flat (in $\nu F_\nu$) for {\it AGILE}
and steep for {\it Fermi}.
We modelled the different 2007 and 2008 states with a different
injected electron distribution, much flatter in 2007,
leaving all other parameters unchanged.
This implied a different number of emitting electrons, and a 
somewhat different jet power (see Tab. \ref{powers}).

\section{Discussion}

Several of our blazars have good {\it Swift} data that allowed to estimate
the black hole mass and the disk luminosity.
There are two reasons for the disk luminosity to emerge
in the optical UV range of the spectrum in these luminous sources:
i) according to the blazar sequence, and directly seen in the {\it Fermi} data,
the high energy spectrum is steep, implying a high energy peak
below the {\it Fermi} energy range. This implies that the corresponding
synchrotron spectrum peaks in the IR band, with a relatively smaller
contribution in the optical--UV.
ii) If highly luminous blazars have large black hole masses, then
we expect relatively smaller magnetic fields in the dissipation region
with respect to blazars with smaller masses (see Ghisellini \& Tavecchio 2009),
and thus an enhanced external Compton component with respect to
the  synchrotron one.
Since ours are the most luminous {\it Fermi} blazars, is likely that
they have black holes with the very large masses.
This is is indeed what we found: all (but one) of them are above $10^9 M_\odot$.
We can compare these values with the ones found, with a different
method, for the quasars found by the Sloan Digital Sky Survey (SDSS)
(see e.g.  Vestergaard \& Osmer 2009; Vestergaard et al. 2008;
Kelly, Vestergaard \& Fan 2008; Natarajan \& Treister 2009;
Shen et al. 2008).
The method to find the black hole mass in this case is to measure
the width of some broad emission line and the luminosity of the optical/UV
continuum.
Since the latter is found to correlate with the typical size of the BLR
(e.g. Kaspi et al. 2007; Bentz et al. 2006), the
hypothesis of virial motion of the broad line clouds allows to estimate the 
black hole mass.
The corresponding uncertainties are large, but studying very large samples
of thousands of quasars it emerged that there is a ``saturation"
value of the black hole mass, around $5\times 10^9$--$10^{10} M_\odot$.
Since our FSRQs are among the most luminous blazars, their black hole masses
are in agreement with what found by the SDSS.
This consistency, found using two different methods for the mass determination,
is reassuring.

We expect that a powerful jet is accompanied by
a close--to--Eddington accretion rate.
This is based on the relation between the jet power and the accretion disk
luminosity, as pointed out in several works. 
The two powers are found to be approximately equal in Rawlings \& Saunders (1991).
More recent work showed that  jets are $\sim$10 times more powerful than the 
luminosity of their disks (CG08; Maraschi \& Tavecchio 2003).

By model fitting, we could derive the intrinsic physical quantities
needed to estimate the power carried by the jet. 
Therefore we confirm the previous hints of a link between the jet power
and the disk luminosity. 
On average, the jet power is larger (but it is of the same order of) 
the disk luminosity, suggesting $P_{\rm j}\approx \dot M c^2$.
For several of our blazars the UVOT/{\it Swift} data allowed a direct 
estimate of the disk luminosity, that together with the black hole mass
estimate give us the disk luminosities in Eddington units:
the values cluster around $L_{\rm d}/L_{\rm Edd} \sim 0.1-0.3$.
We re--iterate that these results are very approximate and must
be taken with caution, given the uncertainties involved (especially in
decomposing the non--thermal and thermal fluxes, and the possibility
to have some residual dust--extinction in the host fame).
One should also take into account the possibility that the {\it observed}
disk luminosity never reaches the Eddington value in standard disks, since
some of the produced radiation may be trapped within the accretion flow
and swallowed by the black hole before it can be radiated away.

Our sample, by construction, includes only very powerful $\gamma$--ray
blazars: for these sources it is more likely that the accretion disk  
directly shows up in the SED.
However, this disk component is seen also in blazar with
not so extreme luminosities as in 3C 273, where it is
always visible, and in PKS 1510--089 (Kataoka et al. 2008) and
3C 279 (Pian et al. 1999).
In Fig. \ref{all} we have marked the positions of these sources
in the $P_{\rm j}$--$L_{\rm d}$ plane.
There might be different reasons for the different ``visibility"
of the accretion disk in blazars.
In 3C 273 its luminosity is of the order of $3\times 10^{46}$ er  s$^{-1}$,
of the same order than the non--thermal observed luminosity.
In Celotti \& Ghisellini (2008) we found that the Doppler boosting
in this source was slightly less than for the rest of the blazars
we analysed.
We therefore suggest that for 3C 273 (a relatively nearby source, $z=0.158$)
we are looking at a jet very slightly misaligned (say $5^\circ$ instead of $3^\circ$),
letting the disk emission to always stick out from 
the non--thermal spectrum.

For the other two sources the accretion disk luminosity is much less than in 3C 273
(i.e. of the order of $10^{45}$ erg s$^{-1}$).
This can be the consequence of a combination of a smaller black hole mass and
a smaller accretion rate with respect to 3C 273, leading also to a somewhat
smaller jet power (see Fig. \ref{all}).

}

Since almost all dissipation regions are located within the BLR,
the most relevant seed photons for scattering are Ly$\alpha$ photons.
As explained in Ghisellini \& Tavecchio (2009), this implies a break
in the intrinsic spectrum at $\sim 15/(1+z)$ GeV
due to the decreasing (with energy) Klein--Nishina cross section
and to the $\gamma$--$\gamma \to e^\pm$ process between high energy
and broad line photons.
If spectra will {\it not} cut--off at $15/(1+z)$ GeV, 
the seed photons must be at lower frequencies (Ghisellini \& Tavecchio 2009),
as occurs if the dissipation region is beyond the BLR, where
the most relevant seed photons becomes the IR photons reprocessed
by the torus.
In this case we also expect larger sizes of the emitting blob
(since it is located at larger distances from the black hole), hence
a relatively slower variability.
As pointed out in Ghisellini \& Tavecchio (2009),
these are also best candidates to study the effect of the 
photon--photon absorption process due to the intergalactic optical--UV
backgrounds.

\section*{Acknowledgments}
We thank the anonymous referee for constructive 
comments.
This work was partly financially supported by a 2007 COFIN--MIUR and
an ASI I/088/06/0) grant.
This research has made use of the NASA/IPAC Extragalactic Database (NED) 
which is operated by the Jet Propulsion Laboratory, California Institute 
of Technology, under contract with the National Aeronautics and Space 
Administration.

\end{document}